\documentclass[final,number,5p,twocolumn]{elsarticle}

\usepackage{geometry}
\usepackage{subfig}
\usepackage{booktabs}
\usepackage{amsmath}
\usepackage{lscape}

\biboptions{}

\makeatletter
  \def\ps@pprintTitle{%
     \let\@oddhead\@empty
     \let\@evenhead\@empty
     \def\@oddfoot{\reset@font\hfil\thepage\hfil}
     \let\@evenfoot\@oddfoot
  }
\makeatother

\journal{Nuclear Instruments and Methods in Physics Research Section A}

\begin{document}

\begin{frontmatter}

\title{A cryostat to hold frozen-spin polarized HD targets in CLAS: HDice-II}

\author[Jlab]{M. M. Lowry\corref{cor1}}
\ead{mlowry@jlab.org}
\author[Jlab]{C. D. Bass \fnref{fn1}}
\author[Jlab,Roma]{A. D'Angelo}
\author[Jlab]{A. Deur}
\author[Jlab]{G. Dezern}
\author[UVA]{C. Hanretty}
\author[Carn]{D. Ho}
\author[Jlab]{T. Kageya}
\author[Jlab]{D. Kashy}
\author[NSU]{M. Khandaker}
\author[Jlab,Cler]{V. Laine}
\author[UConn]{T. O'Connell}
\author[Jlab]{O. Pastor}
\author[UVA]{P. Peng}
\author[Jlab]{A. M. Sandorfi}
\author[Orsay]{D. Sokhan \fnref{fn2}}
\author[Jlab]{X. Wei}
\author[Jlab]{M. Zarecky}

\address[Jlab]{Thomas Jefferson National Accelerator Facility, 12000 Jefferson Av, Newport News, VA 23606, USA}
\address[Roma]{Universita' di Roma \textbackslash Tor Vergata'', and INFN Sezione di Roma \textbackslash Tor Vergata'', Via della Ricerca Scientifica, 1 I-00133 Roma, Italy}
\address[UVA]{University of Virginia, 1400 University Av, Charlottesville, VA 22903, USA}
\address[Carn]{Carnegie-Mellon University, 5000 Forbes Av, Pittsburgh, PA 15213, USA}
\address[NSU]{Norfolk State University, 700 Park Av, Norfolk, VA 23504, USA}
\address[Cler]{Universit\'{e} Blaise Pascal, 34 Avenue Carnot, 63000 Clermont-Ferrand, France}
\address[UConn]{University of Connecticut, 115 N Eagleville Rd, Storrs-Mansfield, CT 06269, USA}
\address[Orsay]{Institut de Physique Nucleaire, Bat 100 - M053, Orsay 91406, France}

\cortext[cor1]{Corresponding author. }
\fntext[fn1]{Present address: Le Moyne College, 1419 Salt Springs Road, Syracuse, NY 13214, USA}
\fntext[fn2]{Present address: University of Glasgow, Glasgow G12 8QQ, Scotland, UK}

\begin{abstract}

The design, fabrication, operation, and performance of a helium-3/4 dilution refrigerator and superconducting magnet system for holding a frozen-spin polarized hydrogen deuteride target in the Jefferson Laboratory CLAS detector during photon beam running is reported. The device operates both vertically (for target loading) and horizontally (for target bombardment).  The device proves capable of maintaining a base temperature of 50 mK and a holding field of 1 Tesla for extended periods.   These characteristics enabled multi-month polarization lifetimes for frozen spin HD targets having proton polarization of up to 50\%  and deuteron up to 27\%.

\end{abstract}


\end{frontmatter}

\pdfoutput=1

\section{Introduction}

This is the second in a series of papers  that describe the apparatus necessary for the condensing, polarizing, handling and bombarding of frozen-spin polarized hydrogen deuteride (HD) targets.  Experiments with polarized targets are essential to unravel the amplitudes in meson photo-production.  First proposed in 1967 \cite{Hon67} and first used in photo-production experiments only recently \cite{Hob09}, this type of polarized target has many attractive features for such measurements, but its operation is considerably more complex than that of a conventional polarized target.  Several commercial cryostats and magnet systems are used in the production process, as well as two specially designed and constructed devices.  The first paper \cite{Bass14} detailed the motivation for the target system, its physics principles, the HD target production and utilization process, and one of the two custom-built cryostats required, the transfer cryostat (TC).  This paper describes the design and operation of the second custom-built cryostat, the in-beam cryostat (IBC), that held polarized HD targets in the center of the CEBAF Large Acceptance Spectrometer (CLAS) for the recently-completed meson photoproduction experiment E06-101 \cite{Ard06} at Jefferson Laboratory.  The in-beam cryostat discussed here is the third generation in a series, and is significantly superior in both long-term reliability and operating temperature.  Discussions of the earlier versions may be found in refs.~\cite{Rig95,Wei04}.  A third paper is in preparation which will deal with the NMR measurement and RF manipulation of the polarization and with the polarization lifetime dependences on temperature, field and concentrations of metastable impurities.

\section{In-beam Cryostat}
\label{ss:IBC}

\subsection{Design Considerations}
\label{ss:design}

\begin{figure}
    \centering
    \includegraphics[width=\linewidth]{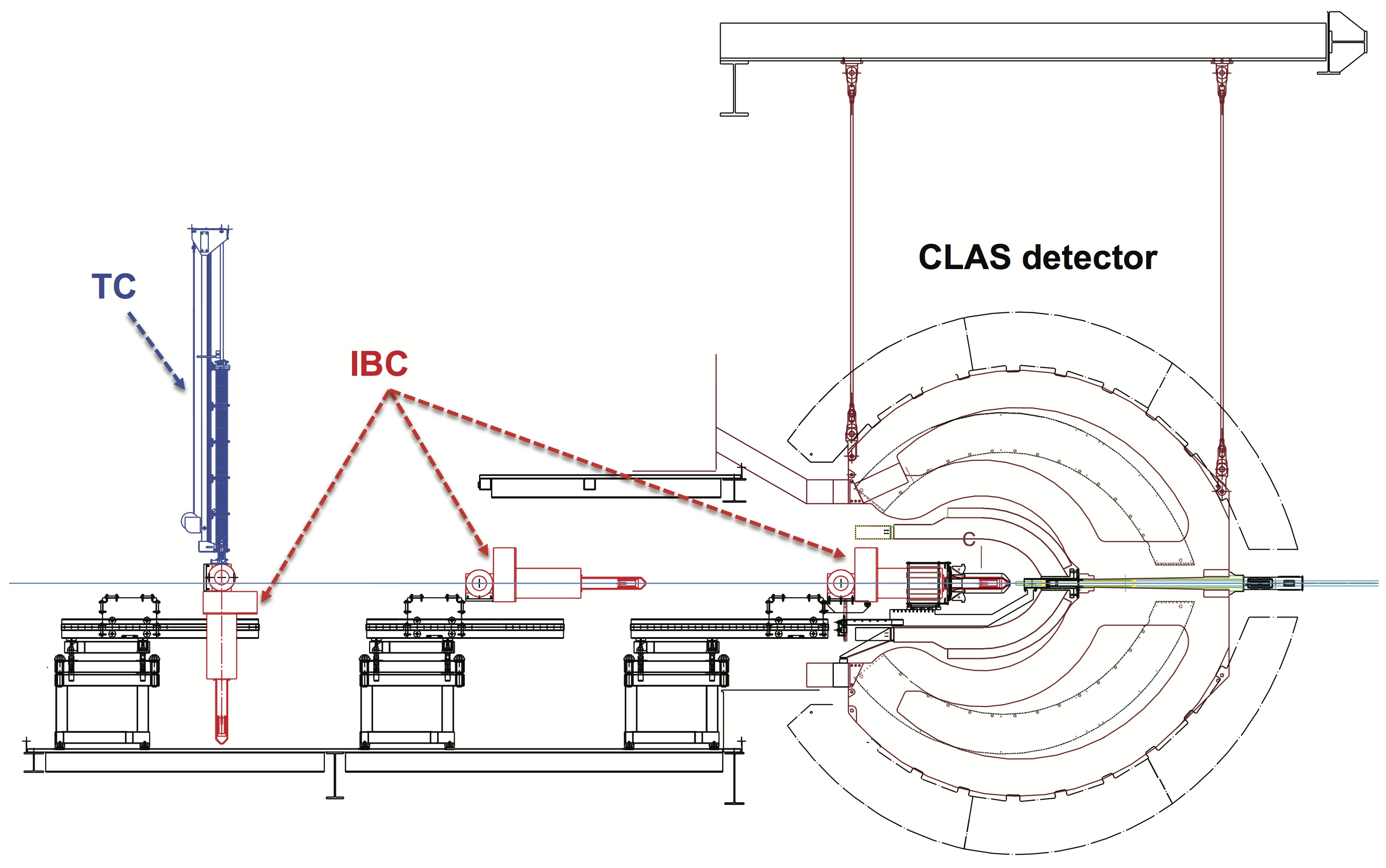}
    \caption{Target manipulations inside Experimental Hall B at Jefferson Lab.  Steps left to right: transfer cryostat (TC - in blue) loads frozen-spin HD target into the in-beam cryostat (IBC - in red); IBC is rotated horizontally; IBC is rolled into CLAS for experiments. (A pumping skid, mounted on the same rails as the IBC, is omitted for clarity.)}
    \label{fig:CLAS}
\end{figure}

The device required to hold frozen-spin HD targets within the CLAS detector at Jefferson Laboratory presents some unique challenges.  The physical dimensions are dictated by the simultaneous requirements of supporting target transfer operations with the TC and holding that target at the center of CLAS.  Since the target transfer is best done vertically while the access to CLAS is horizontal, the refrigerator must operate in both orientations and while traveling between the center of CLAS and a point sufficently distant to allow vertical access to the cryostat (see Figure~\ref{fig:CLAS}). The central access of the cryostat must accomodate the transfer cryostat liquid nitrogen section out to its maximum extension where a docking mechanism must open a 77~K shutter opener on the TC\cite{Bass14} and withstand the 1800 Nt thrust of the transfer cryostat without excessive thermal loading of either cryostat.  Past this shutter opener, the extension of the pumped-liquid-helium TC center tube with target must be accommodated out to the target holder of the IBC.  At the same time, the outer envelope diameter is set by the bore of the CLAS detector and, specifically, the bore of the CLAS Start Counter \cite{Sha06}, and the CLAS detector's large angular acceptance requires thin, low-Z materials to enable escape of particles originating in the target over the range from 0 to 150 degees.

The thermal performance required is dominated by the need to maintain long polarization lifetimes during beam bombardment.  Since a photon beam is a negligible load this translates into a requirement for a base temperature less than 100~mK.   A secondary goal is to maintain a refrigerator temperature under 300~mK during  tests with multi-GeV electron beams, where a 1~nA beam deposits 2.5~mW in the target used for E06-101.  A dilution refrigerator is the only viable option for continuous operation.

The polarization lifetime is not only affected by temperature but also by magnitude of the holding field.  To that end the device requires four magnets of varying maximum strength reflecting the varying lengths of time the target is expected to remain in them and the difficulty in producing that field.  The main holding field is defined as a nominally 1.0~Tesla superconducting solenoid, centered on the target and representing a compromise between lifetime and energy loss of reaction products passing through the coils.  A separate, backup, room temperature solenoid must generate of order 0.01~T in order to guard against main field loss due to equipment failure or operator error.  Between the TC docking point and the main solenoid, a superconducting transfer solenoid with a holding field of at least 0.1~T is needed for transfer operations.  Finally, a superconducting saddle coil must produce a transverse field of at least 0.05~T to allow the holding field to rotate between pointing up and pointing down the beamline.

A final requirement, possibly novel for a dilution refrigerator, is mandated by the U.S. Department of Energy, that pressure vessels meet the requirements of the ASME Boiler and Pressure Vessel Code \cite{BPVC}, which includes vessels with internal or external operating pressures exceeding 103~kPa, and with an inner diameter, width, height, or cross section diagonal greater than 152~mm.  Because of the thin windows for beam entry and exit, this applies to the internal volumes as well as the outer shell.  Satisfying this rule results in significant design complications and costs involving types of materials and thicknesses, joint methods and geometries, and testing regimes.

An immediate consequence is the design decision for all stainless-to-stainless joints to be TIG welded, and all stainless-to-copper and all copper-to-copper joints to be vacuum brazed.  Aluminum-to-stainless joints were commercial friction welds\footnote{Meyer Tool and Mfg., Inc.} and  aluminum-to-aluminum joints were a combination of TIG and e-beam welding.  The exceptions to this are demountable joints which are all indium-sealed.  As fabricated, neither the friction welds nor the e-beam welds proved to be vacuum tight and had to be sealed by painting with epoxy\footnote{Emerson Cummings (Henkel Loctite), Stycast\textregistered 2850-FT catalyst-9}.

\subsection{Central access }
\label{ss:ca}

\begin{figure*}
    \centering
    \includegraphics[width=\linewidth]{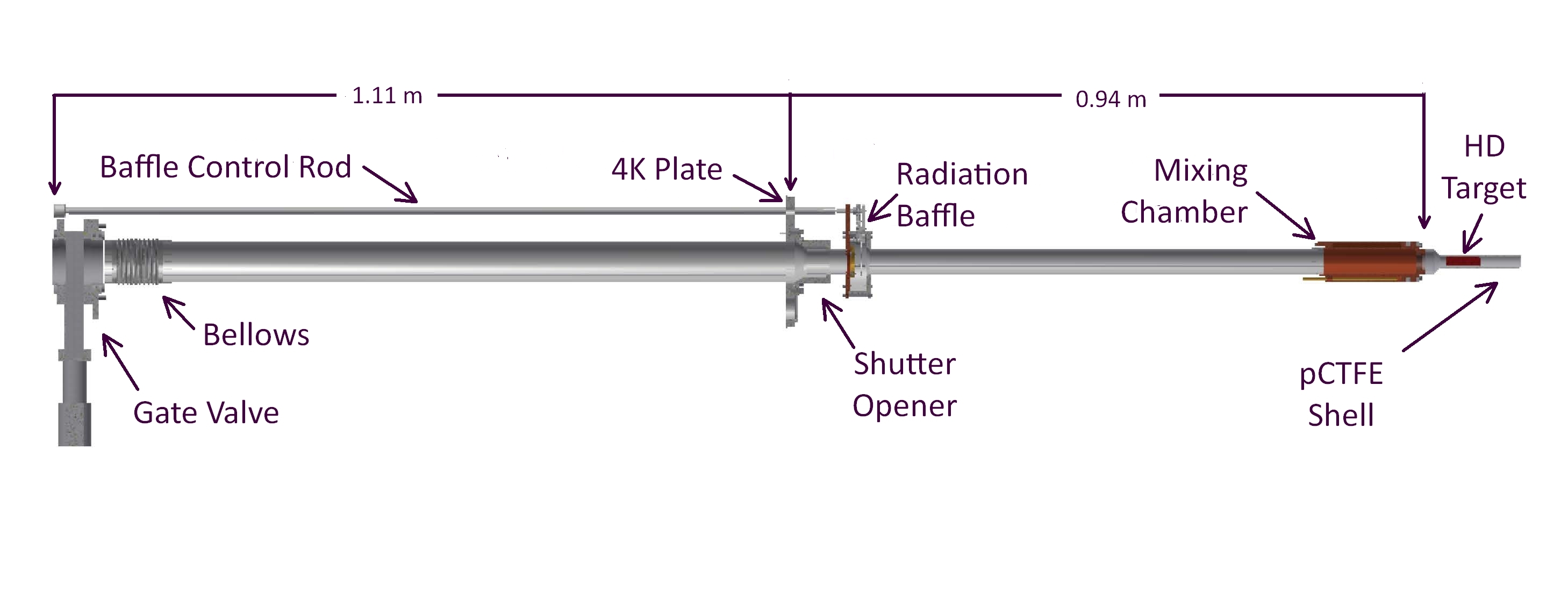}
    \caption{IBC central access cut-away showing gate valve, thermal compensation bellows, shutter opener, radiation baffle, mixing chamber and HD target.}
    \label{fig:ca}
\end{figure*}

  The central access tube is the axis of the design and a separate vacuum space from the main cryostat vacuum, although it is not intended to support a greater differential pressure than about a tenth atmospheric.  It is the path by which HD frozen-spin targets, which are mounted on threaded copper cooling rings \cite{Bass14}, are transferred to and from the target-receiving threads on the mixing chamber.  It is also the beam path to the target. Figure~\ref{fig:ca}  shows a cut-away model view of this assembly.
  
 From the upstream face of the ISO63 gate valve to the mid-plane support plate of the helium-4 reservoir (see Section \ref{ss:LHe}), the access is a 63~mm diameter, 1.11~m long stainless steel tube, with a 102~mm long formed-bellows to relieve thermal stress mounted just downstream of the gate valve. Beyond the 4~K support plate is the demountable shutter opener and downstream of that is the can containing the thermal radiation baffle.

\begin{figure}
    \centering
    \includegraphics[width=\linewidth]{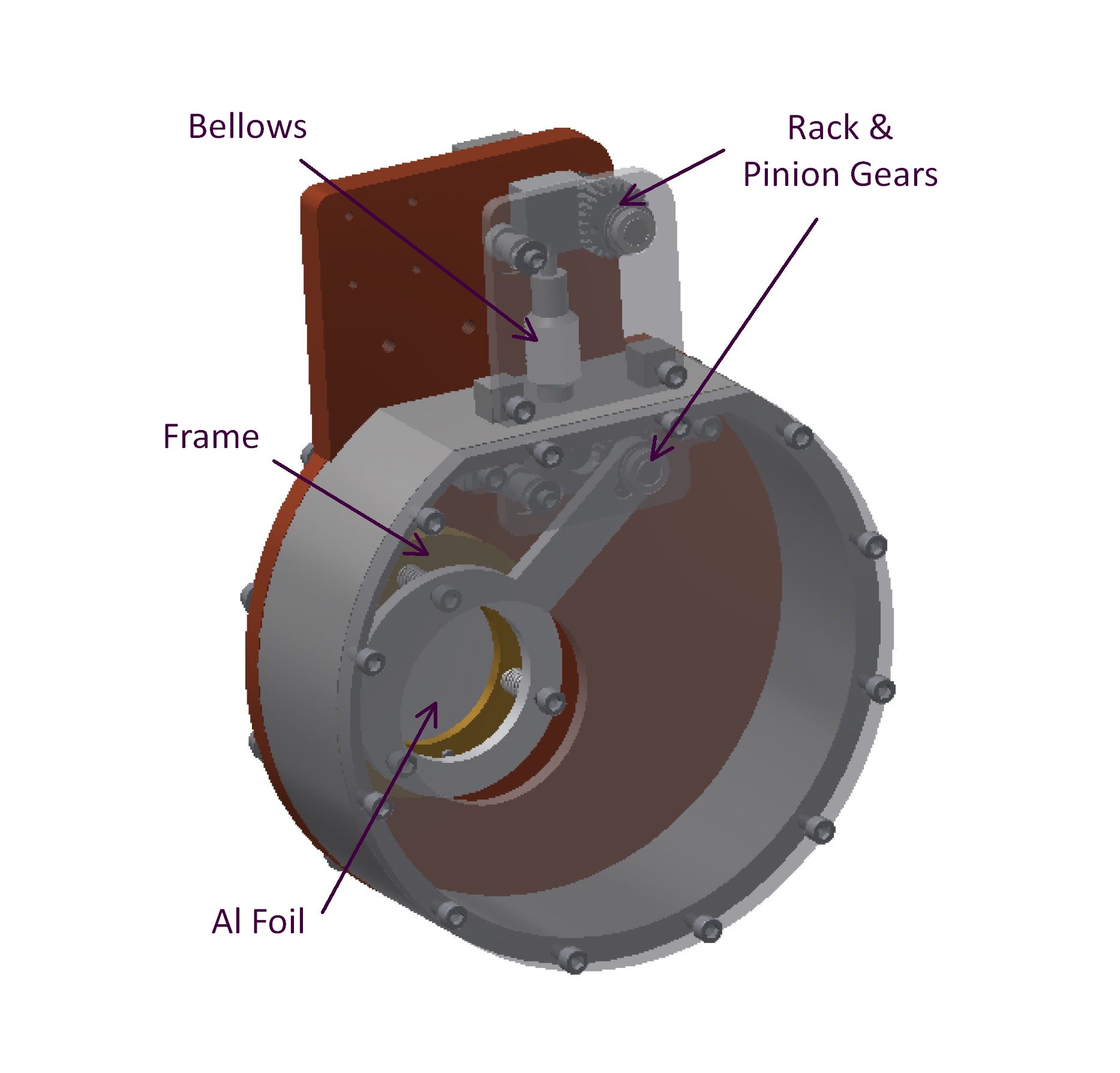}
    \caption{IBC radiation baffle. The metal downstream plates are rendered as transparent in order to show the interior details of the mechanism that moves the swing arm in and out of the beam path.}
    \label{fig:rad}
\end{figure}

    The thermal radiation baffle blocks room-temperature radiation from penetrating to the dilution unit and can be swung out of the way for target transfer (see Figure~\ref{fig:rad}).  The swing is controlled by dual rack-and-pinion gear combinations to translate the rotary motion of the actuator rod to linear motion of a small vacuum sealed bellows and back to the rotary swing of the baffle.  The baffle center is a double layer of 29~mm diameter, 12.5~$\mu$m thick, aluminum foil allowing the photon (or electron) beam to pass with minimal interaction.  The surrounding frame is held in thermal contact with the upstream copper wall of the can by three springs.  Indium seals on the end walls allow access to the interior mechanism. 
    
    Most of the 0.94~m distance from the mid-plane helium-4 plate to the bottom stop of the target holder is occupied by a 38~mm diameter stainless steel tube on which the major components of the helium-3/4 circulation are mounted (see Sections \ref{ss:1K} and \ref{ss:He34}).  This includes the copper mixing chamber vacuum brazed onto the downstream end.  A polyimide spider\footnote{DuPont, Vespel\textregistered-SP1} is mounted on the tube at the point, downstream of the still, where it enters the helium-4 reservoir snout. (It is visible in the right panel of Figure~\ref{fig:1Kpot}.)  It ensures the centering of the tube and removes at least some of the natural droop of such a long horizontal thin tube.
    
    The final 149~mm of the central access is provided by a pCTFE shell just outside the target shell and extending an additional 60~mm downstream.  The downstream end of this shell duplicates the pCTFE thickness of the target cell seen by the beam to that point, and hence the event rate, in a location separable by CLAS from events generated in the target.  This provides a continuous measure of the pCTFE background contribution.

\subsection{Helium-3/4 pumping }
\label{ss:pump}

\begin{figure*}
    \centering
    \includegraphics[width=1\linewidth]{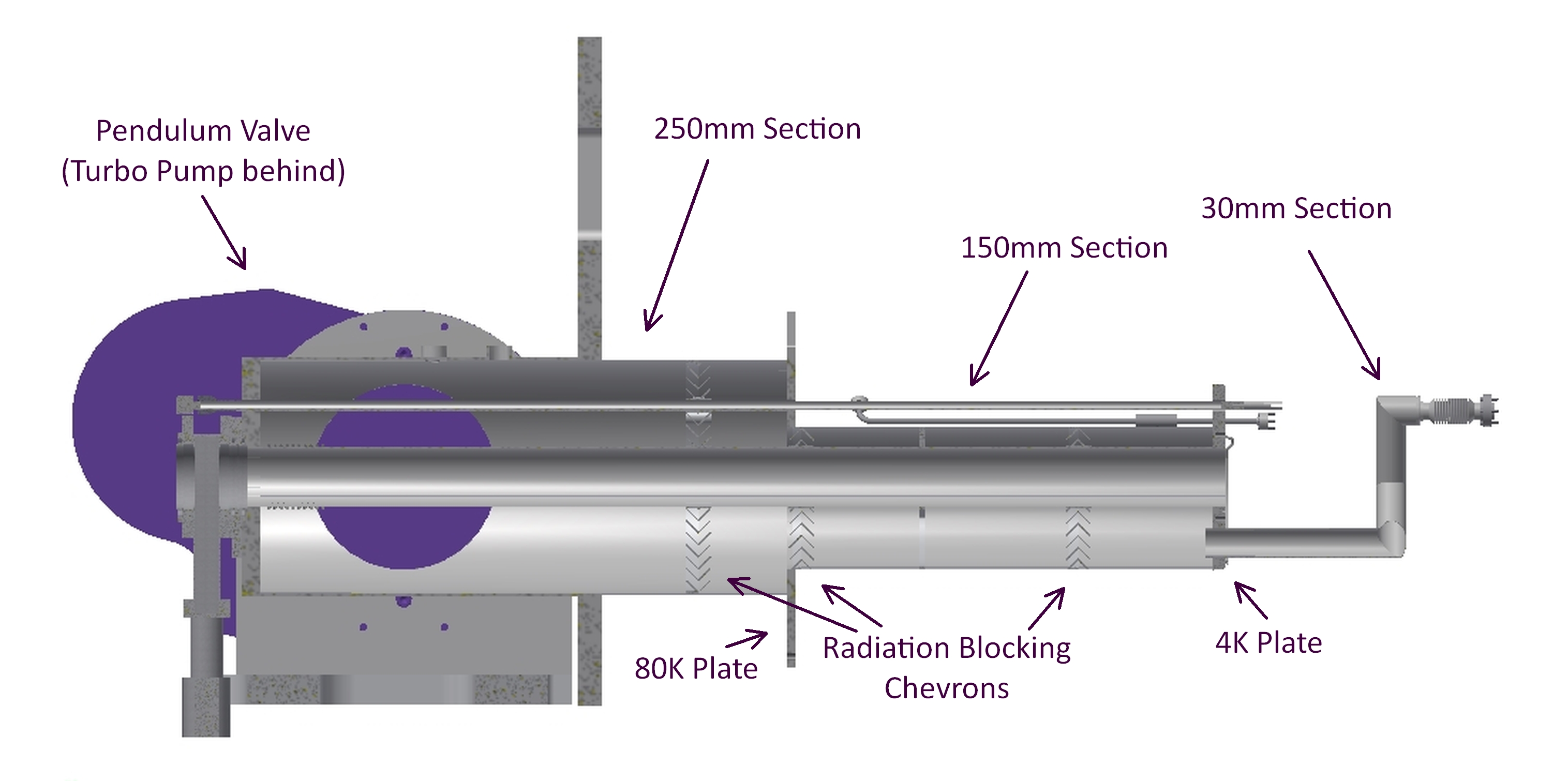}
    \caption{IBC still pumping line.  The steadily increasing diameter maintains pumping speed as the density drops with increasing temperature of the gas.  The chevrons block the direct path for room temperature thermal radiation to reach the still.}
    \label{fig:pump}
\end{figure*}

The primary support structure between the upstream end of the cryostat and the dilution unit (see Section \ref{ss:He34}) is the pump line for the helium-3/4 still.  As shown in Figure~\ref{fig:pump}, the stainless steel line steadily reduces in diameter starting at 250~mm between room temperature and the 80~K plate, down to 150~mm between the 80~K and 4~K plates, and then ranging from 30~mm to 20~mm on either side of the formed bellows connecting to the still.  Three sets of chevron baffles (1 in the 250~mm section and 2 in the 150~mm region) block room temperature radiation from reaching the still. The upstream end is a reducing cross with ISO200 flanges on which are mounted a pair of turbo pumps\footnote{Varian, Turbo-V 1001 Navigator}.  These pumps together have an 1800~ltr/s nominal pumping speed.  This translates to a molar flow of 1.6 milli-moles/s at 2~Pa.  That flow rate implies 1.1~mW of nominal cooling at 0.1~K and 5.4~mW at 0.3~K.  To allow operation at higher still pressures and thus higher potential flow rates, one of the turbo pumps can be removed and replaced with a fixed line to a Roots pump stack\footnote{Pfeiffer, Okta 4000A / 500A} with 1440~ltr/s nominal pumping speed.  Exhausts from all pumps are fed to two scroll pumps\footnote{Edwards, XDS35i} and the gas handling system mounted on the pumping skid located 2-3 meters away (see Section~\ref{ss:skid}).

\subsection{Main helium-4 reservoir }
\label{ss:LHe}

\begin{figure}[ht]
    \centering
    \includegraphics[width=1\linewidth]{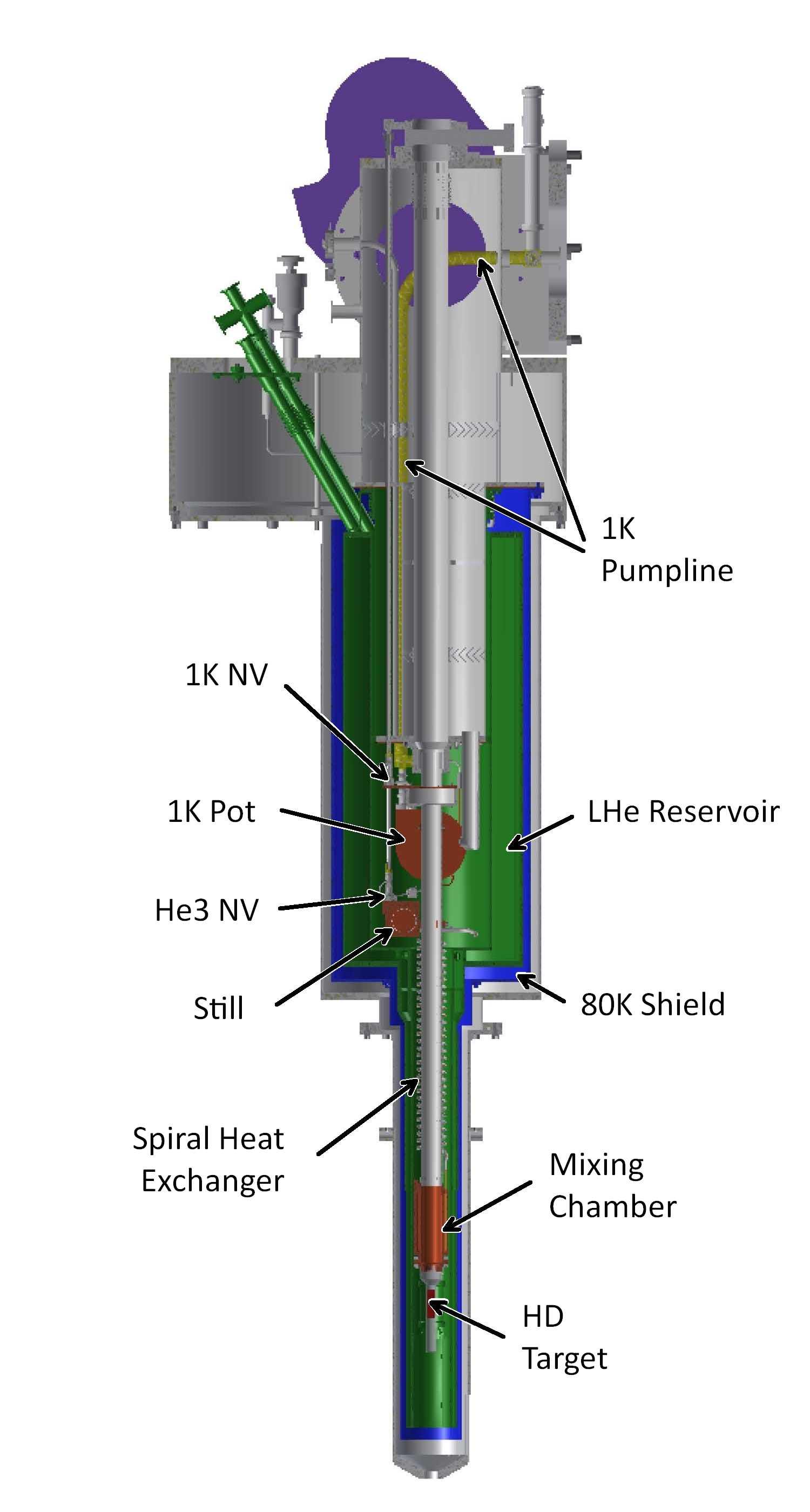}
    \caption{IBC cross section as viewed with the cryostat vertical, as for target loading. An 80~K radiation shield (blue) surrounds the 4~K liquid helium bath (green), which, in turn, surrounds the dilution-cooling components.}
    \label{fig:full}
\end{figure}

The reservoir used to hold liquid helium-4 at a vapor pressure close to atmospheric is two, joined, toroidal cans
(see Figure~\ref{fig:full}).  The upstream stainless steel one, the main volume, has an inner diameter of 212~mm and an outer diameter of 324~mm. The downstream aluminum one, the snout (lower part of Figure~\ref{fig:full}), has diameters of 70~mm and 95~mm, respectively, shrinking to 90~mm near the target.  A thicker region near the base of the snout has the aluminum-to-stainless transitions previously mentioned in Section~\ref{ss:design}. The outside of the reservoir is covered in ten layers of super-insulation and the downstream end of the volume enclosed by the reservoir is capped by a 48~$\mu$m aluminum foil.  The reservoir is shown in green in Figure \ref{fig:full}.

Three of the four magnets described in Section~\ref{ss:Mag} are mounted on the interior surfaces of the inner walls.  Their leads exit through the middle two of the four 25~mm tubes extending at an angle, 30 degree from the axis, at the upstream end (see top of Figure~\ref{fig:full}).  Besides the 3 sets of magnet leads, the tubes carry voltage taps for the normal conductor portion of the leads and wires for two, duplicate, level sensors.  The third tube holds the liquid helium delivery lance and the fourth is the exhaust. The room-temperature plate that is O-ring sealed to the vacuum can and carries the four tubes also has vacuum feedthroughs for the NMR cabling (see Section~\ref{ss:NMR}), which is mounted on the helium can exterior.

Also on the upstream end, at the lowest point when horizontal, a 6.4~mm diameter tube enters and extends to the downstream end of the main can to pickup liquid for cooling the mid-plane helium-4 plate and for transfer to the 1~K pot (see Section~\ref{ss:1K}).   Similarly, another 6.4~mm diameter tube connects at the highest point when horizontal and picks up helium boiloff vapor to cool the 80~K plate and shield (see Section~\ref{ss:LN2}).

\subsection{80~K shield }
\label{ss:LN2}

The boiloff vapor taken from the liquid helium-4 reservoir by the 80~K system flows into a 5.5~turn 220~mm diameter copper coil and then exits the cryostat through a 300~W PID controlled heater.  The gas flow rate is set by a mass flow controller feeding a bellows pump\footnote{Metal Bellows, Model MB602}.  The coil is bolted to the 80~K plate, a copper-clad stainless steel plate making the transition from 150~mm to 250~mm diameter in the helium-3/4 still pumping line (see Figure~\ref{fig:pump} and Section~\ref{ss:pump}).  Bolted to the same plate, is an aluminum thermal radiation shield (blue region of Figure~\ref{fig:full}).  It consists of a 356~mm diameter by 0.87~m long main can and a 108~mm inner diameter by 0.84~m long snout. The diameter of the snount at the base is enlarged to 143~mm for a 76~mm length to accommodate the aluminum to stainless transitions of the helium reservoir snout (see Section~\ref{ss:LHe} and Figure~\ref{fig:full}). The thickness is 3.2~mm except for the final  0.34~m of the downstream end, the target region, where it is thinned to 0.5~mm. The outside of the 80~K shield is covered in ten layers of super-insulation and the downstream end opening is closed by a 48~$\mu$m aluminum foil. 

Platinum thermistors, two located on the coil, one on the main can, and one on the snout, show that with typical helium gas flow rates of 20~ltr/min the coil temperature is about 65~K, the main can about 70~K and the snout about 80~K.

\subsection{4~K plate, 1~K pot and helium-3 return }
\label{ss:1K}

\begin{figure*}
    \begin{center}
    \subfloat{  \begin{minipage}{0.6\linewidth}
                   \includegraphics[width=1\linewidth]{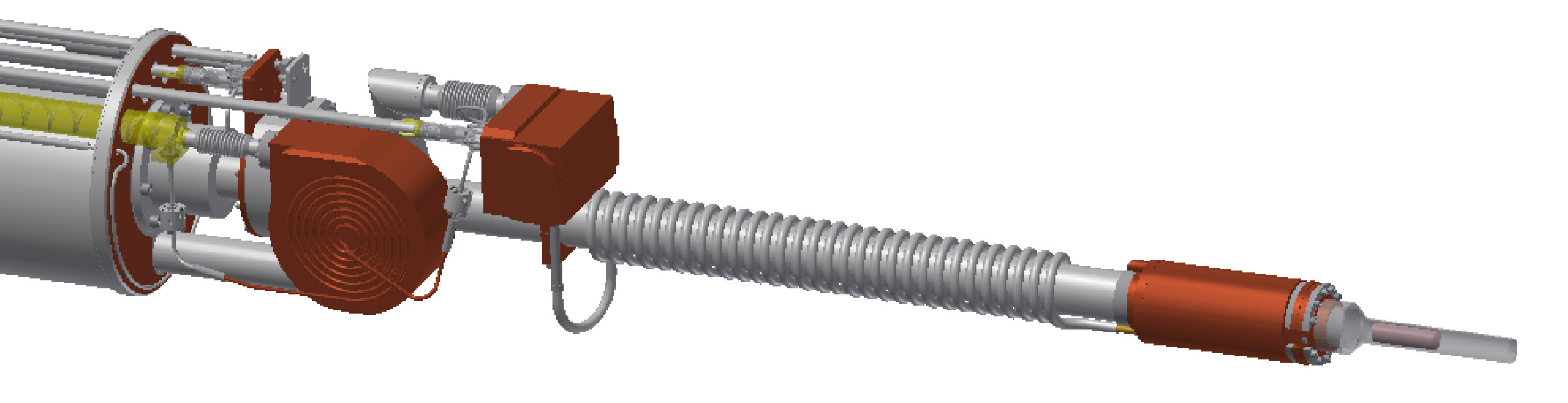}
                   \end{minipage}}
    \\
    \subfloat{  \begin{minipage}{0.45\linewidth}
                   \includegraphics[width=1\linewidth]{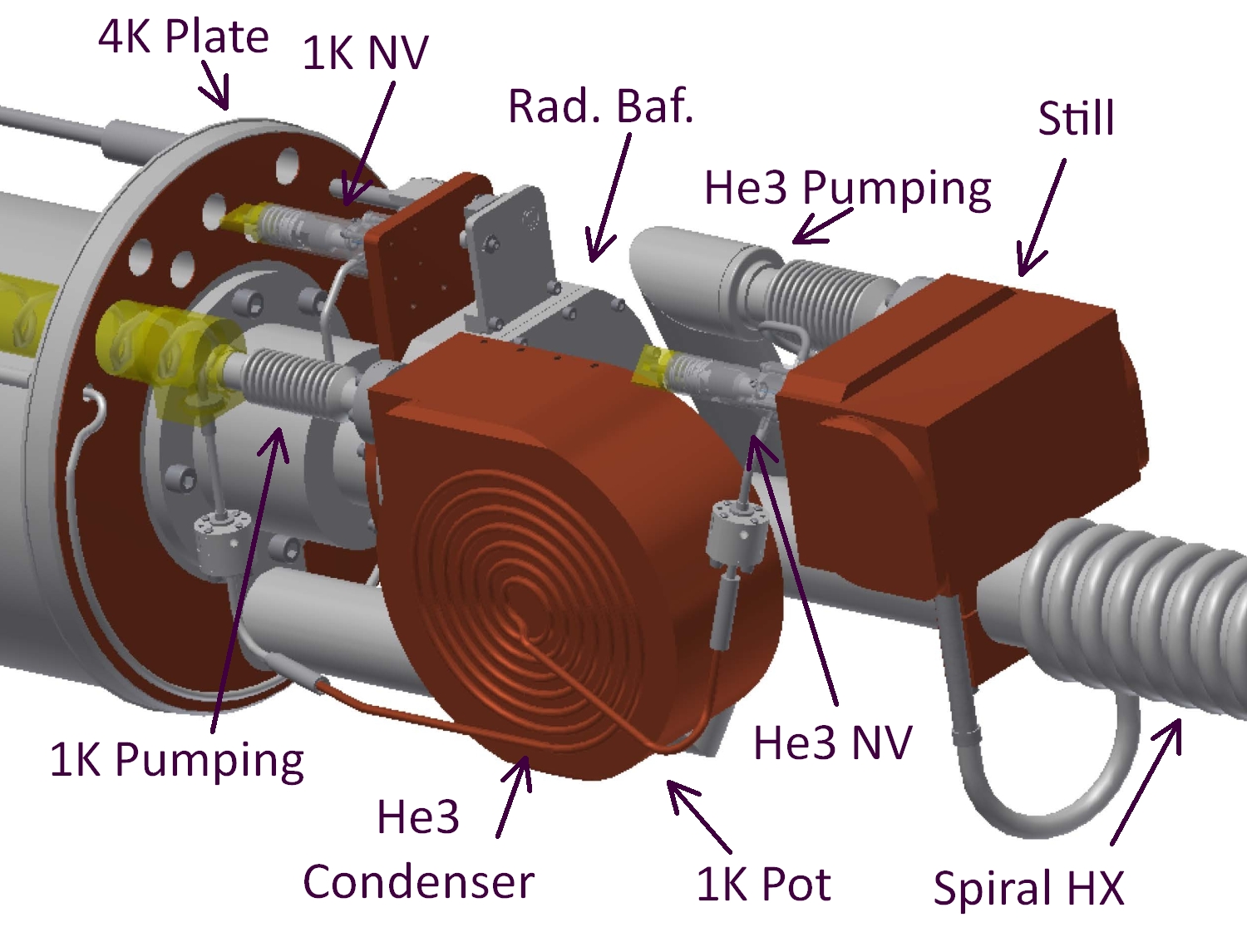}
                   \end{minipage}}
    \hspace{0.02\linewidth}
    \subfloat{  \begin{minipage}{0.45\linewidth}
                   \includegraphics[width=1\linewidth]{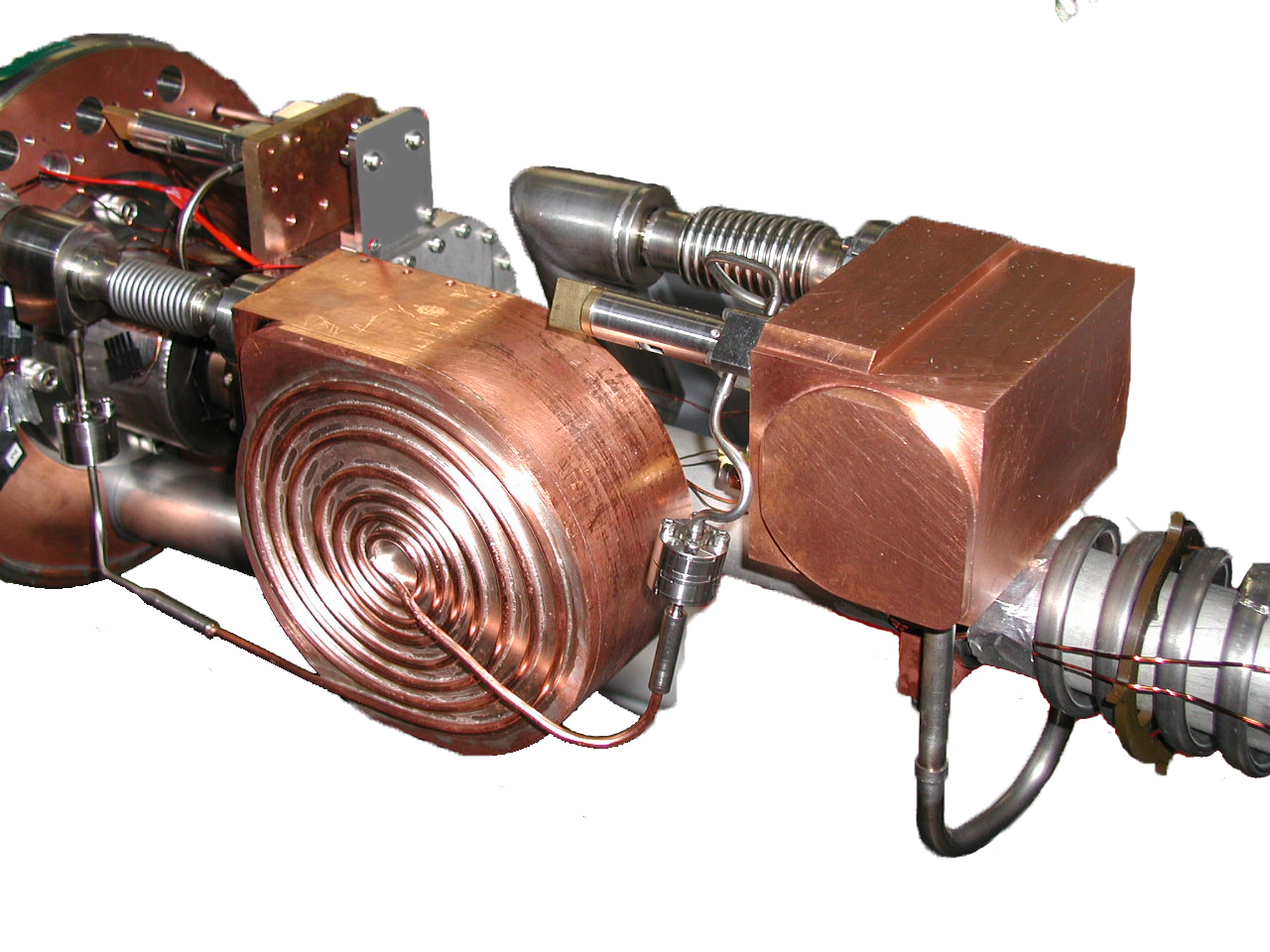}
                   \end{minipage}}
    \caption{Model (above and left) and photograph (right) of the 1~K pot region. The liquid He4 level in the 1~K pot is controlled by its needle (NV), visible at the upper left.  Returning helium-3 enters first through a precooling helical coil within the 1~K pot pumping line (not visible here), next through a spiral condenser line brazed to the 1~K pot and then through a pressure reducing NV mounted on the still. The liquid helium-3 flows through a coil inside the the still and then into the inner tube of the tube-in-tube spiral heat exchanger (HX) shown at lower right.}
    \label{fig:1Kpot}
    \end{center}
\end{figure*}

The plate, referred to as the 4~K plate, that separates the 30~mm to 150~mm transition in the still pumping line (see Section~\ref{ss:pump} and Figure~\ref{fig:pump}) is the principle mechanical support for the liquid helium-4 reservoir (see Section~\ref{ss:LHe}). As shown in Figure~\ref{fig:1Kpot}, this plate also forms the upstream wall of the active cooling portion of the cryostat. In order to assure its temperature a portion of the liquid helium syphoned  from the main reservoir is vaporized in a half-loop of 3~mm copper tubing soldered to the copper cladding of the plate at a rate determined by a mass flow controller. Typical helium gas flow rates of 15~ltr/min keep the plate at about 6~K.

The 1~K pot is a cylindrical volume mounted to the beam-right of the central access, attached to the central access just downstream of the radiation baffle can (see Section~\ref{ss:ca}) and oriented with its axis parallel to the cryostat rotation axis (see Figures~\ref{fig:CLAS}, \ref{fig:full} and \ref{fig:1Kpot}). It has an internal volume of 450~ml and is filled with liquid syphoned from the main reservoir through a needle valve (NV) located on the radiation baffle can (see Section~\ref{ss:NV} and Figure~\ref{fig:1Kpot}).  A pumpline for the 1~K pot parallels the still pumping line ranging in diameter from 12.7~mm at the pot to 25.4~mm at the 4~K plate to 40~mm immediately beyond the cryostat.  Pumps, located on the pumping skid (see Section~\ref{ss:skid}), consist of a nominal 200 ltr/s Roots pump\footnote{Alcatel, RSV 601B} backed by two scroll pumps\footnote{Edwards, XDS35i}.

Returning helium-3 gas is precooled in a 3~meter long, 3.2~mm diameter, stainless tube coiled inside the 1K-pot pumpline from room temperature down to the 4~K plate (see Figure~\ref{fig:1Kpot}). The tube then exits the pumpline and connects to a spiral heat exchanger brazed to the the side of the 1~K pot.  The pumpline heat exchanger proved quite effective, typically delivering 2~K helium-3 to the spiral on the side of the 1-K pot and thereby significantly reducing the load on the 1~K pot.   Finally,  a needle valve mounted on the still provides the flow impedance to maintain the necessary condensing pressure for transforming the gas to liquid. 

\subsection{Dilution unit }
\label{ss:He34}

\begin{figure*}[ht]
    \centering
    \includegraphics[width=\linewidth]{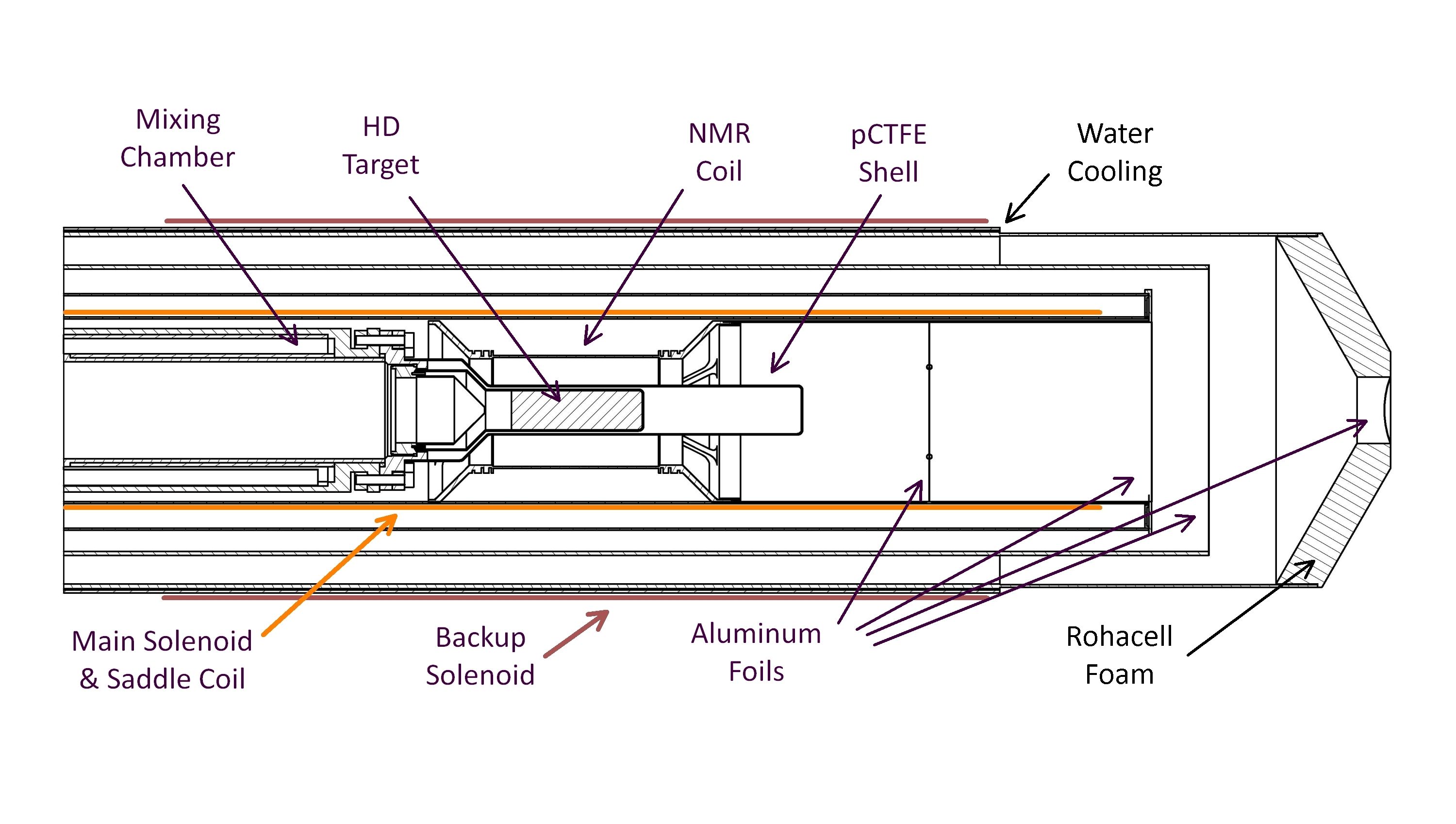}
    \caption{Cross section through the snout of the IBC in the target region.  The section shows the mixing chamber, HD target, NMR coil support, the pCTFE psuedo-target, and the location of various Al foils in the beam path. }
    \label{fig:trg}
\end{figure*}

The still, shown in Figures~\ref{fig:full} and \ref{fig:1Kpot}, is made from an oxygen-free electrolytic (OFE) copper box with a 10.0~cm long by 5.1~cm diameter cylindrical cavity having a tear-drop crosssection (see Figure~\ref{fig:1Kpot}). It is mounted on the top (when horizontal) of the central access with its axis parallel to the cryostat rotation axis(see Figures~\ref{fig:CLAS}, \ref{fig:full} and \ref{fig:1Kpot}).  The tear-drop shape ensures the connection to the spiral heat exchanger is at the low point, independent of cryostat orientation.

  Tubing for the incoming liquid helium-3 forms a heat exchanger inside the still. The 1.2~mm OD by 0.2~mm wall german silver (Cu Ni18 Zn20) capillary is spiral wound six times within the body of the still.  The still temperature is controlled by a heater made from a NiCr thick-film resistor inside a copper slug bolted to the outside of the still.

The capillary continues out of the still into a 7.9~mm OD by 0.5~mm wall stainless tube spiral wrapped about the central access (see Section~\ref{ss:ca}).  Together they form a counterflow, tube-in-tube, spiral heat exchanger (HX) between the helium-3/4 concentrate mixture returning to the mixing chamber and the dilute mixture proceeding to the still. After about 1~meter of linear distance, the inner capillary is enlarged to a 2.0~mm OD by 0.25~mm wall german silver capillary.  After an additional 2~meters, two 300~mm lengths of 2.4~mm OD by 0.36~mm wall copper tube are inserted.  The copper tubes are sintered inside and out with 0.5~mm thick layers of copper powder and separated from each other by 150~mm of the 2.0~mm capillary.  A final length of the 2.0~mm capillary emerges inside the mixing chamber from a 6.3~mm diameter extension of the 7.9~mm diameter tube.  This is at the downstream end of the mixing chamber and at the bottom when in horizontal orientation.  The returning concentrate then bubbles up through the heavier dilute to the top of the mixing chamber. This injection of the return concentrate below the phase boundary is a distinctive feature of the design.  Dissolved helium-3 in the dilute mixture is driven up the 6.3~mm extension, through the 7.9~mm tube and into the still, by osmotic pressure. There it is distilled from the mixture by the applied still heat and pumped away, eventually returning as concentrate to complete the cooling cycle.

The mixing chamber volume (Figure~\ref{fig:trg}) is constructed from two OFE copper cylindrical shells, an outer 62.5~mm OD by 2.3~mm wall shell extending up from the downstream end plate and an inner 40.1~mm OD by 1.5~mm one extending down from the upstream end.   Layers of copper powder 1.5~mm thick are sintered to the interior surfaces of the copper shells.  The inside length of the volume is 103~mm.  On the downstream end of the mixing chamber a copper target holding ring with the necessary M35x1 thread is bolted with an indium seal both to preserve the vacuum integrity of the central access and to ensure good thermal contact. Two calibrated ruthenium oxide RTD thermometers\footnote{Lake Shore Cryotronics, RX-102A and RX-102B} are mounted on the ring.  The accuracy of the calibrations is $\pm 4$~mK at 50~mK.  The resistance is measured by an AC resistance bridge\footnote{PicoWatt, AVS-47B} on the pumping skid (see Section~\ref{ss:skid}) with a pre-amp immediately adjacent to the cryostat.

\subsection{Needle valves }
\label{ss:NV}

The in-beam cryostat requires two needle valves.  One controls the flow to the 1K-pot and the other serves as a variable condensing restriction in the helium-3/4 circuit.  The requirements for such valves are simple to state:  1) No leaks to the outside, although vacuum tight shut off is not required. 2) Low thermal leaks to room temperature and 3) good flow control of low viscosity cryogenic liquids in the low flow regime. Additional considerations are low dead volume and small size. The numerical flow requirements are given in Table~\ref{tbl:NV}.

\begin{table}[ht]
\caption{Needle valve flow requirements.}
\centering
\begin{tabular}{ c c c c }
  \toprule
  liquid    & flow range & viscosity             & pressure \\
            &  [liter/hr]   & [10$^{-6}$ Pa s] & [bar] \\
  \midrule
  $^{4}$He    & 0.2 -- 2.0        & 3.3 (4~K)                  & 1.0 \\
  $^{3}$He    & 0.01 -- 1.0       & 2.9 (1~K)                  & 0.1 \\
  \bottomrule
\end{tabular}
\label{tbl:NV}
\end{table}

\begin{figure*}
    \centering
    \subfloat{\includegraphics[width=0.30\linewidth]{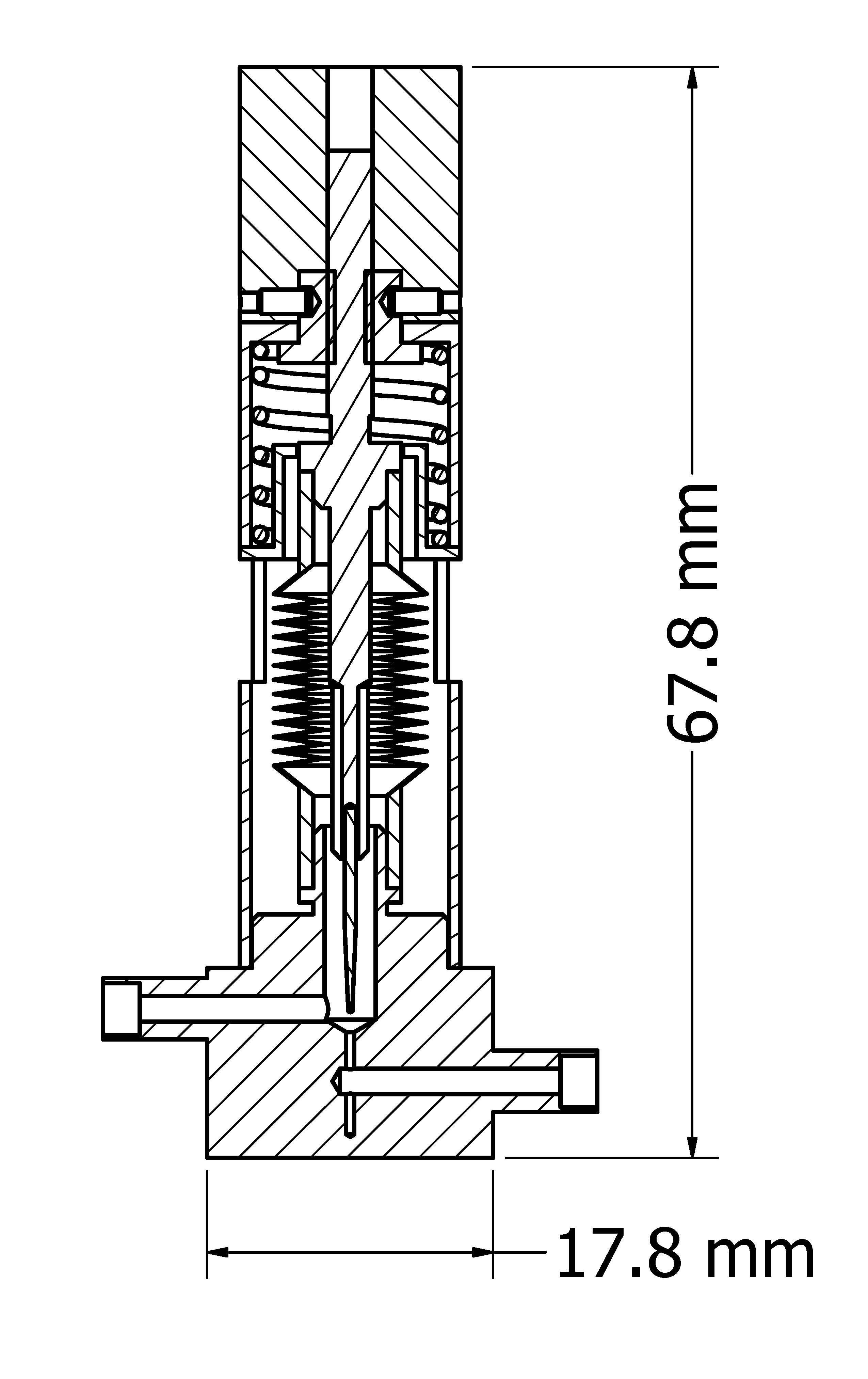}}
    \subfloat{\includegraphics[width=0.70\linewidth]{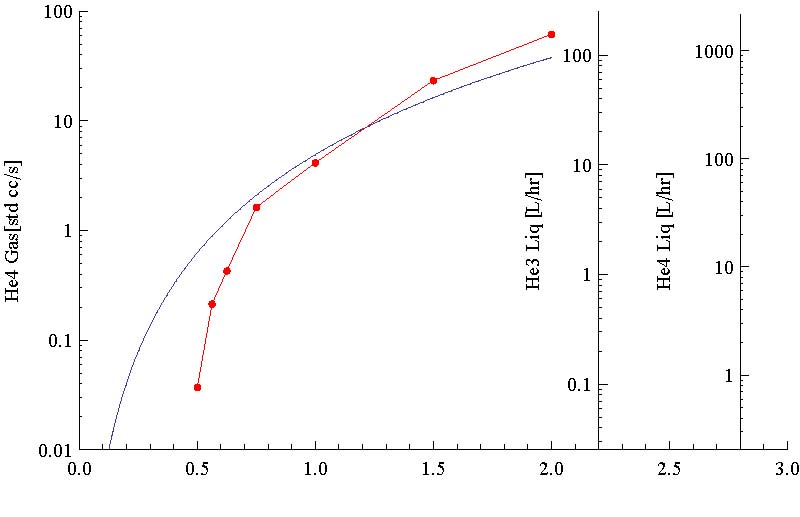}}
    \caption{
Design and performance of needle valves used to control flow to the 1~K pot and pressure in the helium-3 condenser. Calculated (blue curve) and measured (red points) flows for helium-4 gas at 300~K and 1.0~bar differential pressure are plotted against the valve opening in numbers of turns of the \#3-56 UNF thread that positions the tungsten-carbide needle.  The additional scales to the right shown implied flows for helium-3 liquid at 1~K and 0.1~bar and helium-4 liquid at 4~K and 1.0~bar.
 }
    \label{fig:NV}
\end{figure*}

A design was developed to meet these needs. A cross section is shown in Figure~\ref{fig:NV}. A tungsten carbide needle\footnote{Semprex Corporation, p/n 31-5071.} with a diameter of 0.7~mm and a tip angle of $5.0^{\circ}$ is vacuum brazed into one end of a stainless steel stem.  The other end is threaded and  positions the needle part way into a 0.5 mm cylindrical hole in the stainless steel base.  Base and stem are connected by a 9.5~mm OD welded bellows\footnote{Standard Bellows Company, p/n 37-12-2-EE.}.   A cover holds the nut for the thread, the connection to a control rod, and a spring\footnote{Associated Spring, p/n C0480-038-1250-S.} to suppress backlash. A guide to prevent bellows rotation and limit travel pokes through slots in the cover. The design is shown in Figure~\ref{fig:NV} along with a plot of volume flow versus turns open showing experimental results in red from tests with room temperature He4 gas and a theoretical prediction in blue based on a simplified geometry.  Additional scales give the flows for the two cryogenic liquids, helium-3 at 1~K and 0.1~bar and helium-4 at 4~K and 1.0~bar. The plot shows the middle of the required flow range occurs at about half a turn open and also depicts the large flows available for the initial cool down. The needle valves (NV) are visible in Figure~\ref{fig:1Kpot}, both in the photograph and in the model of the cryostat interior.

\subsection{Magnets }
\label{ss:Mag}

As shown in Figure~\ref{fig:trg}, a 4K-liquid-helium-volume surrounds the target region and contains the superconducting magnets that are cooled by direct contact with liquid (see Section~\ref{ss:LHe}).  Outside of this is an 80K-shield of aluminium that is indirectly cooled by contact with evaporating liquid helium vapor at the upstream end of the IBC.  Beyond this is the vacuum can with a water-cooled jacket and the room-temperature backup solenoid wound on that jacket.

In the liquid helium reservoir are the main solenoid, the saddle coil, and the transfer solenoid that are each wound from Supercon 54S43 superconducting wire.  This multistrand wire consists of 54 filaments of NbTi embedded within a copper stabilizer with a ratio of $ 1.0:1.3 $ of NbTi to copper.  The diameter of each multistrand wire bundle is 0.229~mm and an additional layer of varnish insulator increases the overall diameter to 0.254~mm.

The room-temperature-to-helium transitions for all three sets of magnet leads are an update of the helium vapor cooled design of K. R. Efferson\cite{Eff67}. The leads for the three magnets utilize two high current, 8-pin feedthroughs mounted on 2.75 Del Seal flanges\footnote{Insulator Seal (MDC Vacuum Products), p/n 9142010}.  These feedthroughs are rated for 23~A per pin but the air-side connector is only rated for 17~A per pin. On the main reservoir side, each pin connects to a 1.6~mm diameter copper braid inside a 2.2~mm inner diameter pTFE tube. The braid is soldered to the superconducting wire at the bottom, downstream end of the large diameter portion of the main reservoir. One pair of pins is used by the transfer magnet, three pairs for the main solenoid and four pairs for the saddle coil. Voltage taps are made at room temperature and at the normal-to-superconducting joint on each lead of the three magnets. This gives three voltage differences for each magnet; two monitor the voltage drop on the normal portions and the third monitors the superconducting status of the coil.  Two flow controllers, one on each 8-pin feedthrough, are adjusted manually to set the helium vapor flow from the reservoir to the helium return system through the two groups of pTFE tubes on the basis of the voltage drops in the normal sections of the leads.

\begin{figure}[ht]
    \centering
    \includegraphics[width=\linewidth]{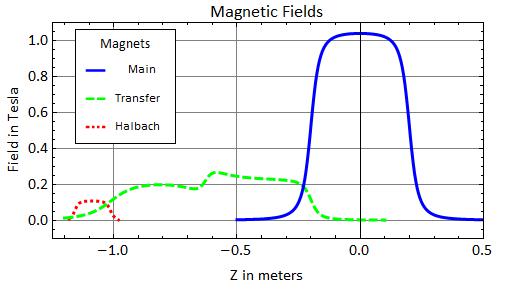}
    \caption{Plot of the magnitude of the fields on axis generated by two of the IBC magnets as a function of distance from the target position. The main solenoid is in blue and the transfer solenoid in dashed green.  The TC Halbach permanent magnet array \cite{Bass14} is shown in dotted red at the docking position to illustrate the holding fields during target loading.  The rapid change in the transfer solenoid field at about $Z=-0.6$ is the transition between the large diameter main bath portion and the small diameter snout portion. The other two IBC magnets are not plotted.  The saddle coil is not used during target loading and would only just be visible at this scale (0.075~T).  The backup coil is too weak to show up (0.014~T).}
    \label{fig:mplot}
\end{figure}

The transfer magnet is two double-layer solenoids in series, one extending from the center of the liquid helium reservoir to the downstream end of the large diameter, a 400~mm length with a 220~mm diameter, and one from the base of the snout to within 200~mm of the target center, a 420~mm length, at 72~mm diameter.  When powered to 17~A during the transfer of a polarized HD target, as shown in Figure~\ref{fig:mplot}, the coil provides a $>0.15$ T holding field over the full travel distance of the target from the transfer cryostat's Halbach magnet docked in the opener (see Figure~\ref{fig:ca}) to the main solenoid magnet of the in-beam cryostat.  Current is supplied by a manually controlled power supply\footnote{Cryomagnetics, Model CS-4}.

\begin{figure}[ht]
    \centering
    \includegraphics[width=\linewidth]{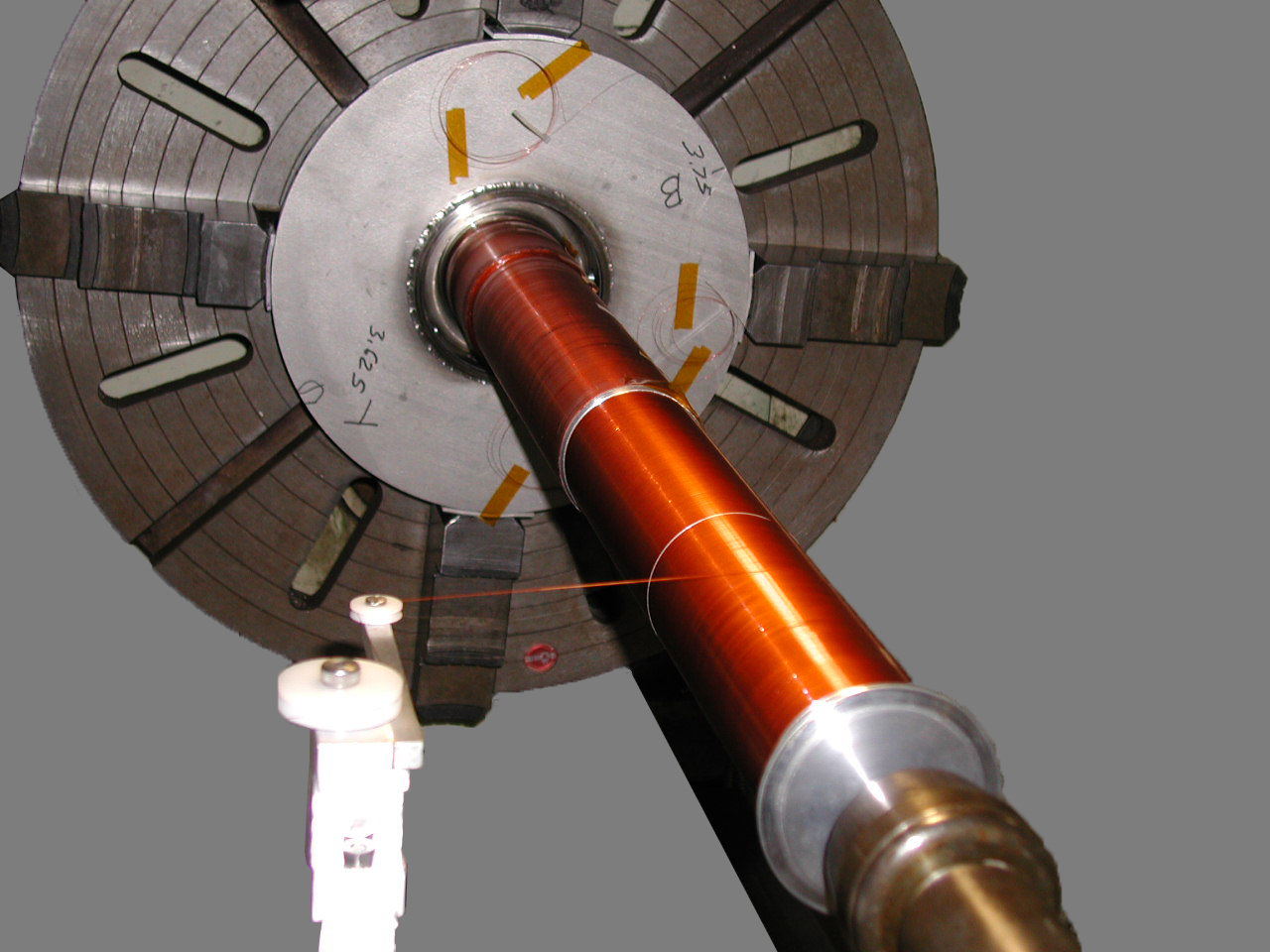}
    \caption{IBC main solenoid during winding of the second of four layers. Also visible upstream of it is the already wound snout portion of the transfer solenoid.  Just visible in the the middle of the main solenoid is a gap of one wire width created by an aluminum fin that is only two layers high.  The reduction in turn density improves field uniformity.}
    \label{fig:mag}
\end{figure}

Both the transfer magnet and the main solenoid were fabricated on a lathe-based winding machine, as shown in Figure~\ref{fig:mag}.  The main solenoid has a total length of 400~mm and shares the same center as the HD.  This solenoid is wound from four layers of wire and has an inner diameter of 71.32~mm.  Voids between wire bundles are filled with Stycast 1265 epoxy\footnote{Emerson Cummings (Henkel Loctite), Stycast\textregistered~1265.}.  
A one-wire width, two-layer deep gap in the center gives a design uniformity over the target volume of $10^{-4}$ but fabrication imperfections may limit the actual uniformity.  The main solenoid produces a nominal field of 1~T at a current of 51~A, as shown in Figure~\ref{fig:mplot}, and is driven by a power supply\footnote{Oxford Instruments, Intelligent Power Supply IPS120-20} controlled by software on the NMR computer system.

A saddle coil is placed around the main solenoid and also centered on the target.  The coil is used to temporarily hold the target polarization transverse to the cryostat axis while the direction of the main solenoid is reversed, thus reversing both the H and D polarizations.
The saddle coil is made from a pair of single layer, 54-turn coils that are wrapped around the cylindrical nose of the IBC to form a bedstead coil geometry.  The pair is 150~mm long at a diameter of  73.58~mm and encased in a 0.25~mm layer of Stycast 1265 epoxy.  The saddle coil generates a field of 0.075~T at a current of 60~A and is driven by a manually controlled power supply\footnote{Oxford Instruments, IPS120-20}.

A non-superconducting, room temperature, backup solenoid with a diameter of 290~mm and a length of 280~mm is also centered on the HD target.  The backup solenoid is wound on a water cooled substrate integral to the snout vacuum can as indicated in Figure~\ref{fig:trg}. It has three layers of wire, 1.02~mm diameter aluminum plus 50~$\mu$m thick insulation.  The backup solenoid produces a field of 0.014~T at a current of 9~A supplied by a manually controlled power supply\footnote{Kepco, BOP 50-20MG}.

 Also apparent in Figure~\ref{fig:trg} is the relatively open geometry for viewing particles emitted from the target over a polar angle from 0 degrees back to 150 degrees. All the cylindrical shells outside the target are either thin aluminum or pCTFE.  The only higher Z material is the 4 layers of superconducting wire in the main solenoid  and the single layer in the saddle coil (the 3 layers of wire in the backup solenoid are aluminum). The seemingly thick material supporting the downstream aluminum exit window is a low-density, high-strength, closed-cell foam\footnote{Evonik Industries, Rohacell 110 WF}. 
 
\subsection{NMR }
\label{ss:NMR}

\begin{figure*}
    \centering
    \subfloat{  \begin{minipage}{0.60\linewidth}
                   \includegraphics[width=1\linewidth]{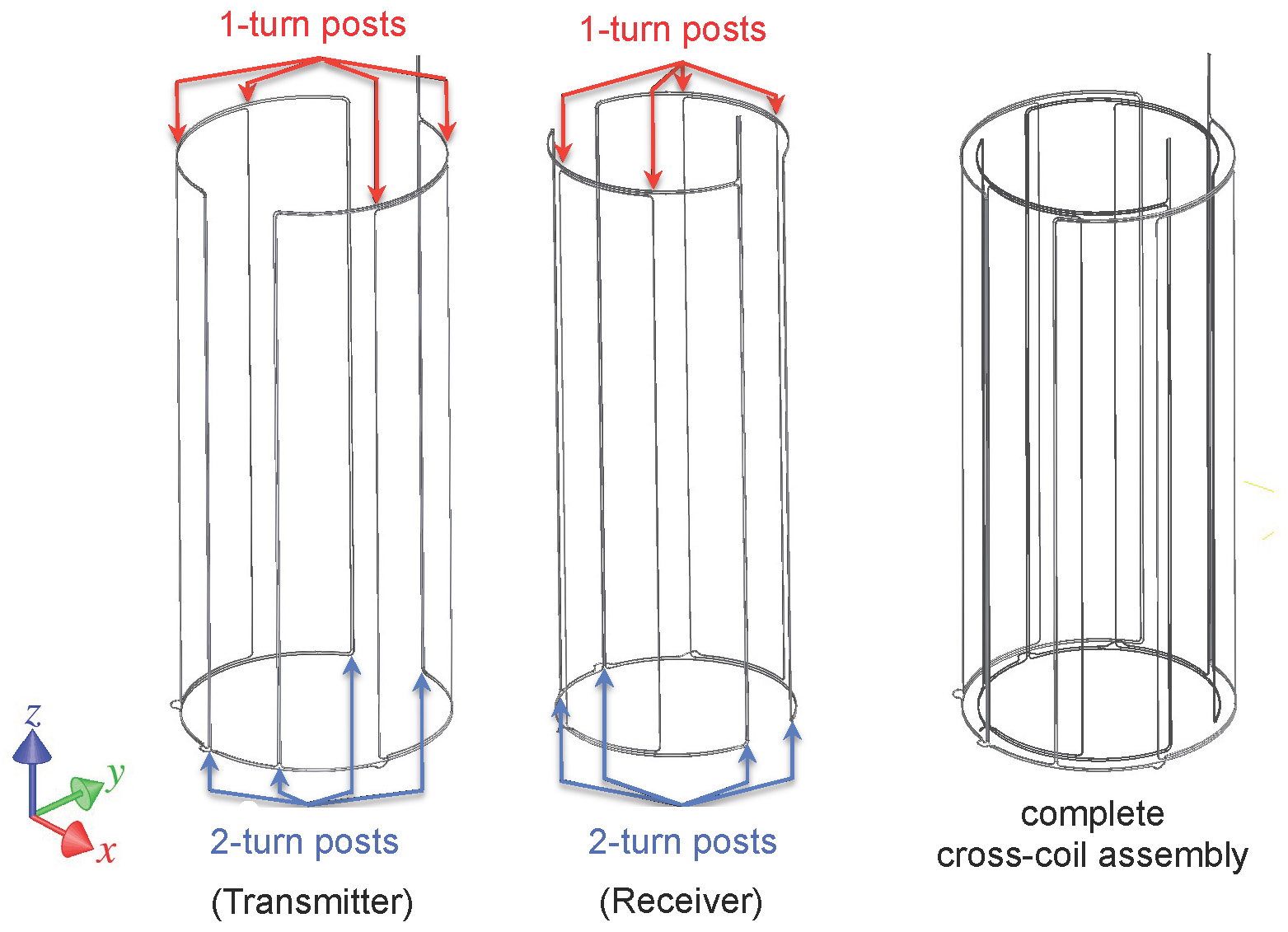}
                   \end{minipage}}
    \hfill
    \subfloat{  \begin{minipage}{0.30\linewidth}
                   \includegraphics[width=1\linewidth]{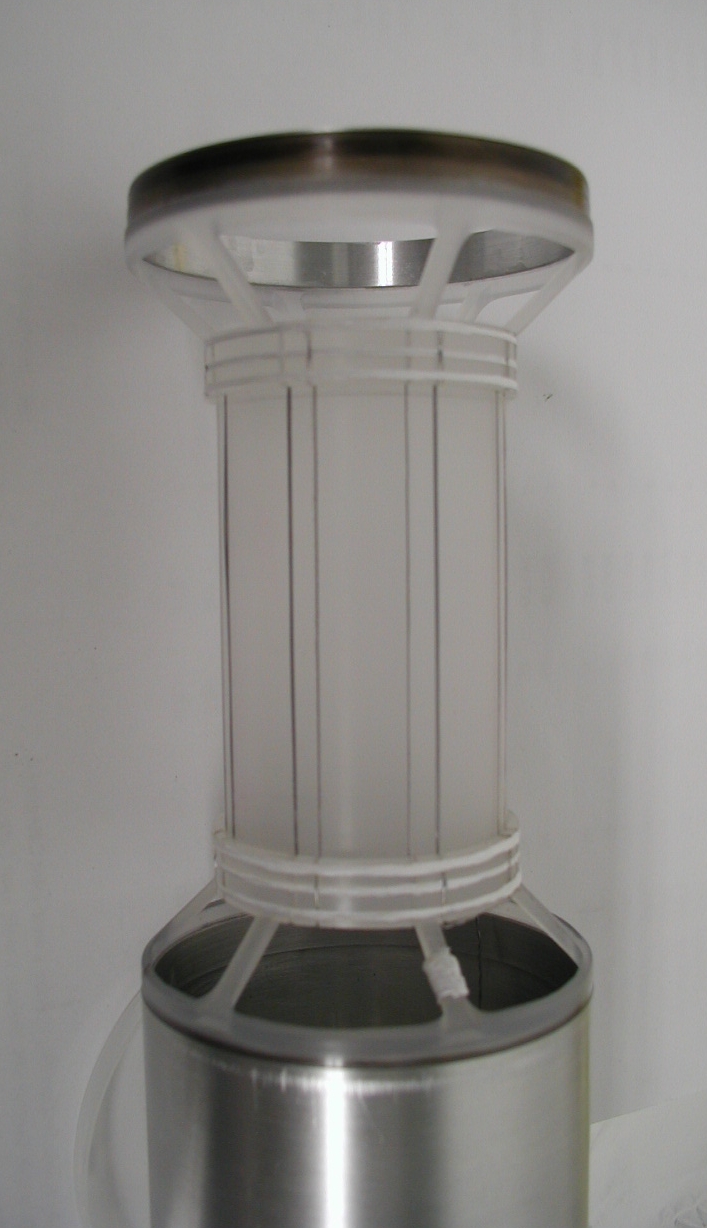}
                   \end{minipage}}
    \caption{NMR coil wiring plan and photograph of completed pair of crossed coils. Each coil is composed of 6 turns, three on either side of the target.  Two of the turns on a side subtend an angle of 144 degrees and the third an angle of 56.8 degrees. The assembly approximates a ``bird cage'' geometry and improves the uniformity of the RF field \cite{Wei13}. }
    \label{fig:NMRcoil}
\end{figure*}

\begin{figure}[ht]
    \centering
  \subfloat{  \begin{minipage}{0.48\linewidth}
                   \includegraphics[width=1\linewidth]{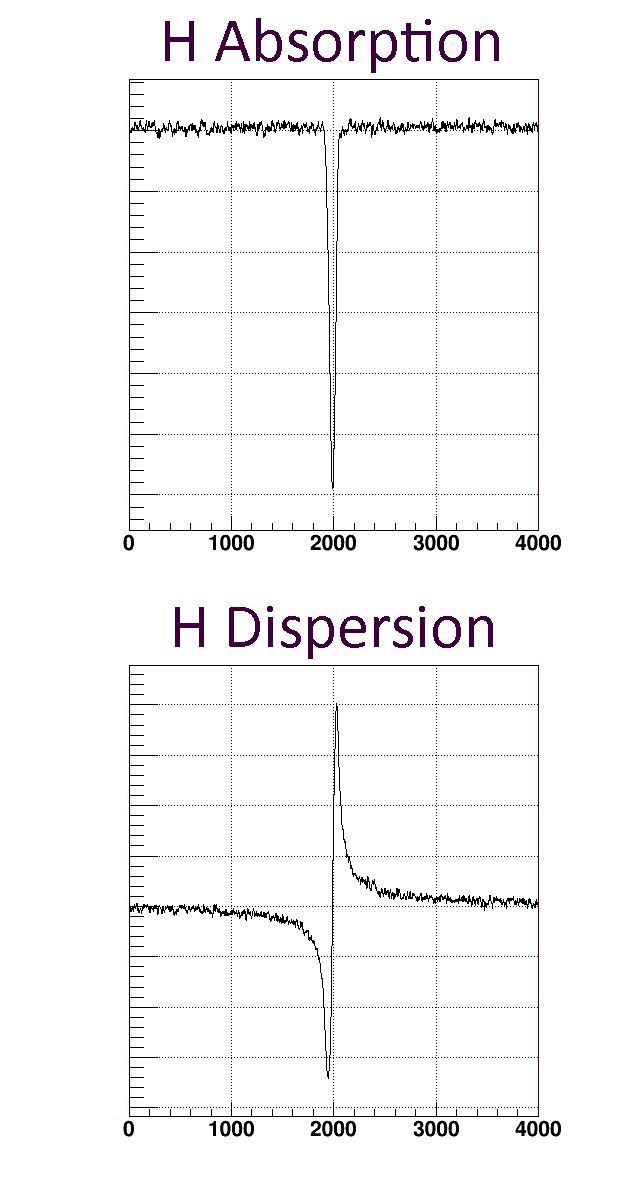}
                   \end{minipage}}
    \hfill
    \subfloat{  \begin{minipage}{0.48\linewidth}
                   \includegraphics[width=1\linewidth]{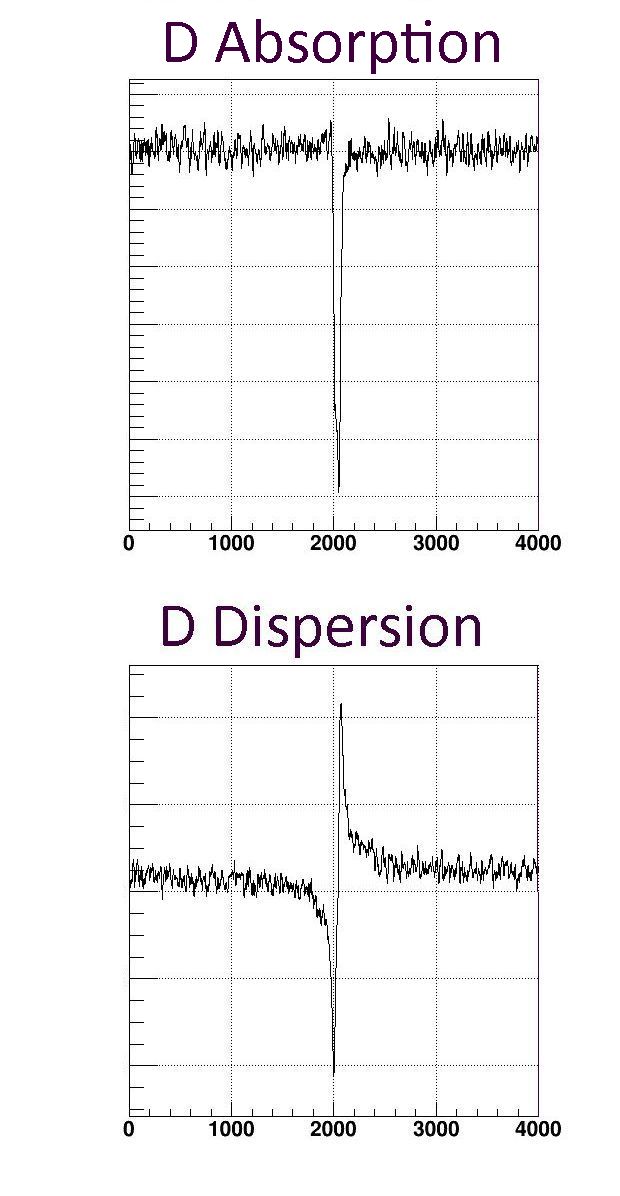}
                   \end{minipage}}
    \caption{Typical absorption and dispersion signals for H and D generated by frozen-spin polarized HD targets as seen with the IBC NMR coils. The RF frequency is held constant (11.15~MHz for H, 1.757~MHz for D) and the magnetic field is ramped.  The vertical axis is response in arbitrary units and the horizontal axis is data point number (0.076 Gauss per point).  Full length of the trace is 300 Gauss.  The D line is not split by the deuteron electric quadrupole moment because the HD crystal symmetry results in no electric field gradient.}
    \label{fig:NMRsig}
\end{figure}

The target polarization is monitored and manipulated by radio frequency electromagnetic interactions generated and received by coils that are wound on a 41.7~mm diameter pCTFE mandrel surrounding the target.  The mandrel is supported by an array of 2.5~mm thick pCTFE ribs (see Figures~\ref{fig:trg} and \ref{fig:NMRcoil}) connecting it to a pair of 69.6~mm diameter, coaxial pCTFE hoops .  The ribs are oriented to sit in the shadow of the CLAS torus coils to minimize their impact on particle energy loss.  The downstream hoop is attached to a 69.6~mm diameter, 0.5~mm thick aluminum support tube that is a slide fit within the liquid helium snout. At the end of the snout, the coil leads are connected to coax cables that lead to the vacuum feedthroughs on the helium reservoir exhaust plate (see Section~\ref{ss:LHe}).  The coil approximates a ``bird cage'' geometry, with a cosine current variation, in order to improve the RF field uniformity. The coil assembly is shown in Figure~\ref{fig:NMRcoil}.  Figure~\ref{fig:NMRsig} shows typical absorption and dispersion signals obtained obtained with these coils for both protons and deuterons.   Further information on the construction, theory and results of these coils, and on the computer-controlled system that transmits and receives radio-frequency NMR signals with them, can be found in Reference~\cite{Wei13}.

\subsection{Pumping skid}
\label{ss:skid}

\begin{figure}
    \centering
    \includegraphics[width=\linewidth]{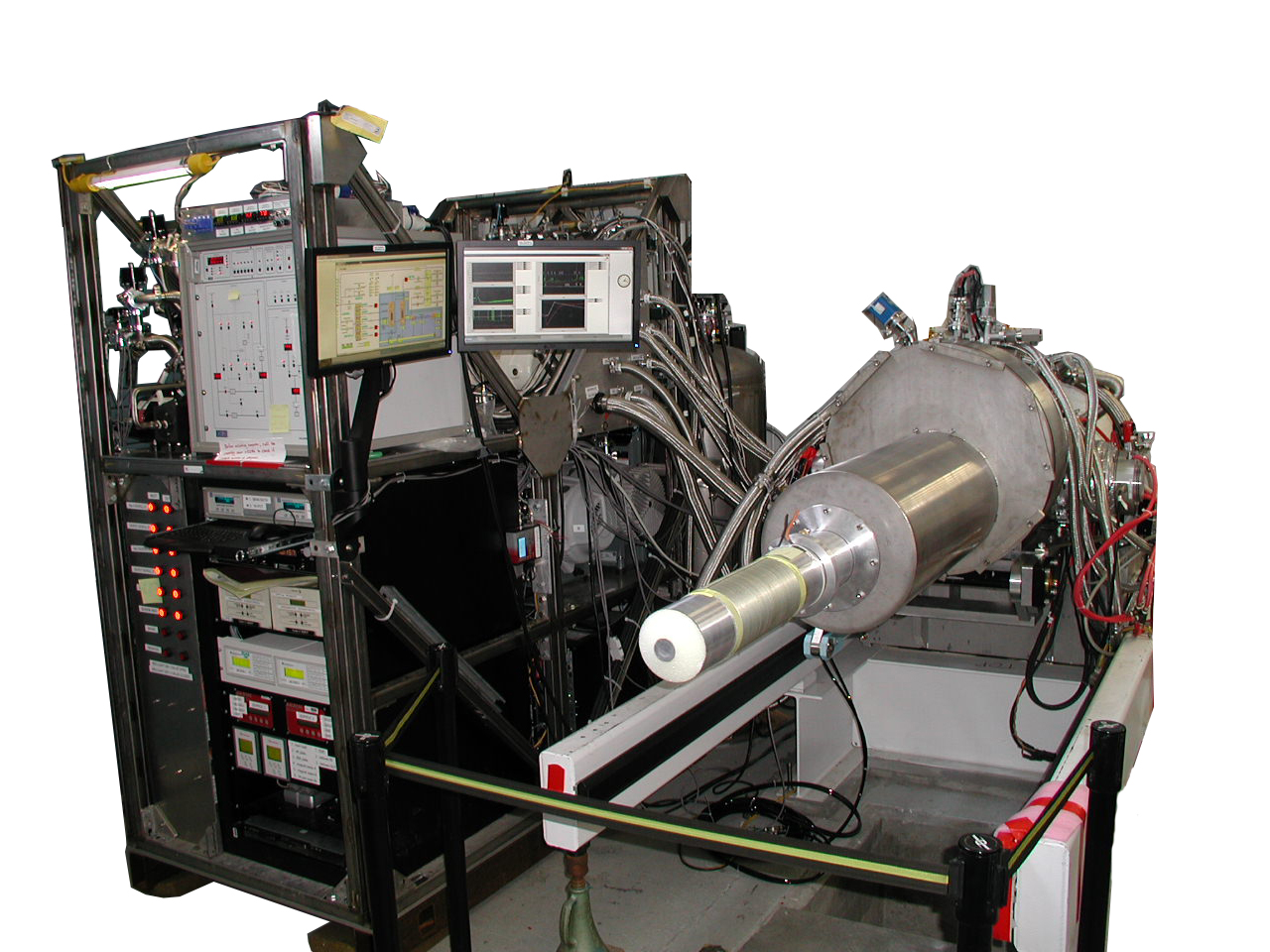}
    \caption{IBC pumping skid and cryostat during initial tests. The aluminum wire of the backup solenoid (see Section \ref{ss:Mag}) is visible between the two yellow bands on the IBC snout.}
    \label{fig:skid}
\end{figure}

The pumping skid (see Figure \ref{fig:skid}) provides a single support structure for the many supplementary pumps, flow controllers, valves, heaters, and gauges needed by the multiple gas flows generated by the cryostat.  Two  sealed pumps\footnote{Metal Bellows, MB602} (170~ltr/min each) move exhaust helium gas into the helium gas recovery system from five flow controllers: one on the main reservoir, one on the 4K plate, one on the 80 K coil, and one on each of the two magnet lead multipin feedthroughs.  Also mounted on the skid are the two scroll pumps\footnote{Edwards, XDS35i} backing the 1~K pot Roots pump\footnote{Alcatel, 601B}.  They exhaust into the recovery system directly.

The helium-3/4 gas handling is done by a computer-interfaced gas-handling system\footnote{Oxford Instruments, Intelligent Gas Handler}. Augmenting it are a backing pump subsystem of two scroll pumps\footnote{Edwards, XDS35i}, an impurity trapping subsystem of two coldtraps in an automatically re-filled liquid nitrogen dewar and a 280~liter dump tank storage subsystem (sufficient for the mixture of 50~liters of helium-3 and 185~liters of helium-4) .  All the components are mounted on the pumping skid.

In addition, the readout and control electronics are mounted on the skid.  The three temperaure sensors operating below 0.5~K are readout by a multi-channel AC resistance bridge\footnote{PicoWatt, AVS-47B}, the other twelve are readout by two multi-sensor temperature monitors\footnote{LakeShore Cryotronics, Model 218E}.  Diaphragm gauges and two instrument controllers\footnote{Edwards Vacuum, ASG and TIC} monitor pressures at nine points in the gas recovery system. Two control modules\footnote{Aalborg, SDPROC} handle the six flow controllers.  Three level sensor readouts\footnote{Cryomagnetics, Model LM-510} monitor the levels in the main reservoir, the 1~K pot, and the liquid nitrogen trap which has an autofill module as well.  A controller\footnote{National Instruments, RealTime and CompactRIO} provides remote operation of gas system valves and the scroll pumps. All the electronics and pumps are powered through four uninterruptible power supplies mounted on the skid as well.  These can power the IBC systems for upto ten minutes, providing more than sufficient time for a backup diesel generator to start and reach full speed.

\subsection{Computer control system}
\label{ss:pc}

 All of the skid-mounted control and monitoring elements (see Section \ref{ss:skid}), in turn, interface to a rack mounted computer that reads them out and sets parameters. This was done through a program, written in LabVIEW\footnote{National Instruments}, that displays values on a system schematic and allows parameter values to be adjusted.  The computer program also feeds the information into an EPICS\footnote{LANL-ANL Collaboration, Open Source software} data stream for offline storage and use by the accelerator control and cryogenic plant, and transfers the data into a local utility for periodic logging and into a local strip chart utility to allow monitoring of trends and recent history.

\section{Operation and Performance}
\subsection{Cooldown}

The refrigerator is cooled down from room temperature while the cryostat is vertical.  The liquid helium transfer lance is placed into an inital fill cone that directs the flow down to the base of the snout and liquid helium is transferred.  The boil-off cools the magnet leads, the 1K-pot, the 4K-plate, the 80K-plate and, if a vapor-cooled transfer line is used, the transfer line itself.  It takes from one to three days for the main reservoir to cool enough to begin accumulating liquid helium, depending on the transfer rate, ranging from 15~ltr/hr to 1.5~ltr/hr.  Once the snout is filled, the main reservoir level will begin to indicate and the 1K-pot will begin to fill.  At that point the delivery lance is lifted out of the inital fill cone and the level in the reservoir rises to the nominal operating level of 40-60\%.
A small amount of helium-3/4 gas mixture is circulated (2-3 ltr/min) in the dilution unit to cool it.  More mixture is required as the unit cools until eventually the entire inventory is condensed in.  This process takes one to three days.  Then power is applied to the still to maintain circulation and the refrigerator cools to base temperature in 6 to 12 hours.

\subsection{Target transfer }

\begin{figure}
    \centering
    \includegraphics[width=\linewidth]{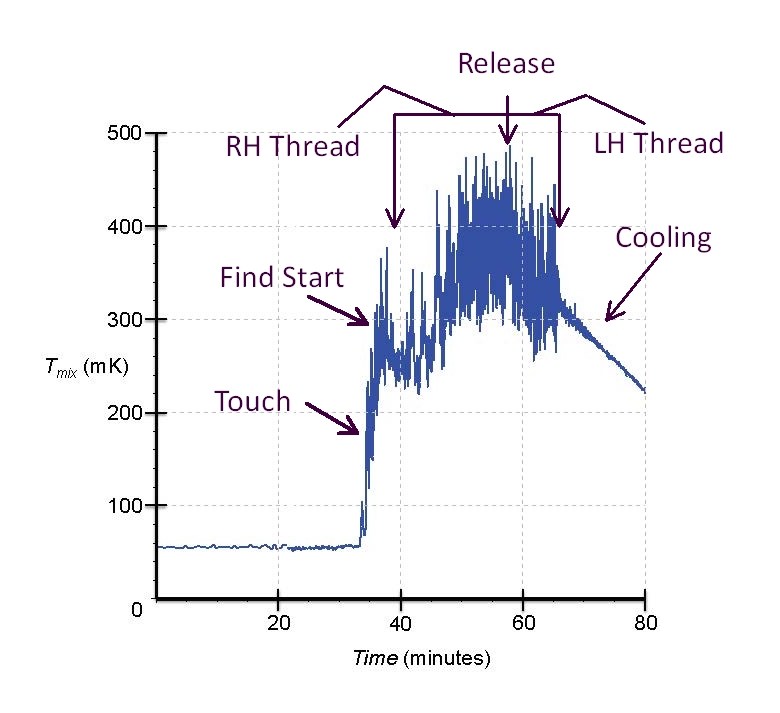}
    \caption{IBC mixing chamber temperature during a typical target transfer, in this case, the loading of target~19b. The varous stages of the transfer: 1) touch of the TC on the mixing chamber, 2) aligning of the thread starts, 3) threading in of the right-hand (RH) outer threads on the target ring, 4) release and un-threading of the left-hand (LH) inner threads, and 5) the beginning of return to base temperature, are marked.}
    \label{fig:xfer}
\end{figure}

As shown in Figure \ref{fig:CLAS}, the transfer cryostat (TC) can be attached to the in-beam cryostat (IBC) while the IBC is vertical.  The joining spool piece is evacuated, gate valves are opened and the vacuum of the TC is joined to the centeral access of the IBC.  As described in Reference~\cite{Bass14}, the liquid nitrogen temperature portion of the TC can be lowered down the central access to mate the TC thermal radiation shutter with its opener on the 4~K plate of the IBC.  This also seals the central access to prevent room temperature radiation from shining down the central access.  Thus the IBC radiation baffle can be opened without significantly loading the dilution unit.  The loading increases as the 2~K center tube of the TC is lowered through the unit to touch the target-holding threaded ring on the mixing chamber, typically, producing a temperature of 0.2 to 0.4~K.  Clockwise rotation (if a target is to be inserted, counterclockwise if one is to be removed) attaches the target to the mixing chamber (or TC) and then detaches it from the the TC (or mixing chamber). Mechanical energy is dissipated on the unit and along with the thermal contact to the 2~K TC typically produces a temperature increase to about 0.4~K to 0.6~K but the helium-3/4 flow rate does not markedly increase.  The refrigerator starts to cool back down as soon as the threads disengage and continues to do so as the TC center tube is raised back into the liquid nitrogen section, the IBC baffle is closed and the TC is withdrawn from the central access.  The IBC typically returns to base in 3 to 6~hours.

  These features can be seen in Figure~\ref{fig:xfer} which charts the mixing chamber temperature history during a typical target transfer (loading of target 19b; for subsequent polarization history of this target see Figure~\ref{fig:polar}).  After the initial touch of the 2~K target ring on the 60~mK holder, the target is rotated counterclockwise until the starts of the two threads are aligned, producing a perceptible 1~mm drop.  A clockwise rotation then screws in the outer, right-hand (RH) threads, steadily increasing the thermal contact between the TC and the IBC and generating frictional heating spikes.  When the RH threads bottom out, further rotation releases and unscrews the inner,  left-hand (LH) threads, steadily lowering the thermal contact but producing yet more frictional spikes.  When the LH threads decouple, the IBC continues to cool and the spikes cease. 

\subsection{Rotation}

In order to place the target in the beam, the cryostat must be rotated to its horizontal position.  There are two areas of concern in this process.  One is the upstream wall of the helium reservoir.  Because the reservoir is stainless steel, a temperature difference exists between the wall areas in contact with liquid and those only in contact with vapor.  A slow rotation is thus needed to avoid a sudden boiloff and rapid pressure rise that could interfere with liquid helium delivery and possibly uncover the magnet.  The other area of concern is the 1~K pot pumpline.  If the pot has been allowed to overfill, there can be liquid in the bottom of the pumpline which will flow up the line just as the cryostat reaches horizontal.  This will rapidly cool the return helium-3 capillary, condensing the gas and collapsing the return pressure.  When the 1~K helium-4 evaporates and the cooling stops, the helium-3 will vaporize in the small volume capillary generating an over pressure that can stop circulation and force mixture into the dump.  This can be avoided by care in setting the 1~K pot level before executing the procedure.

Following bombardment, the cryostat must return to vertical in order to remove the expended target.  This process is easier than the rotation above because the two issues mentioned are not a problem in this direction.

\subsection{Photon and Electron Beam Running}

A photon beam on target produces no discernable effect on the cryostat beyond the necessity to operate it with a networked computer running Remote Desktop.  The story is quite different for electron beam bombardment.  The multi-GeV, minimum-ionizing beam deposits about 2.5~mW/nA in a 50~mm long target \cite{Lowry13}.  Heating from a nano-Amp beam thus requires running at maximum circulation rate and still produces mixing chamber temperatures in the 150 to 250~mK range.

\subsection{Warmup}

Once the target is transferred out, the cryostat can be warmed to room temperature.  The first step is to recover the helium-3 portion of the mixture, which takes about 2 hours.  The larger and higher latent heat helium-4 portion takes another 12 to 24 hours, depending on heater power used.  At this stage, the delivery of liquid helium to the main bath is stopped and the reservoir empties in a few hours.  The cryostat then warms to near room temperature in about two days.  

\section{Discussion}

\begin{figure}
    \centering
    \includegraphics[width=\linewidth]{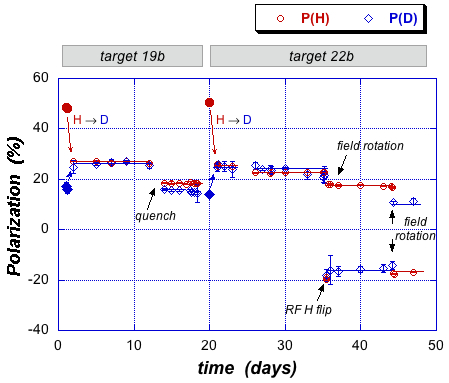}
    \caption{Polarization history of two of the HD targets during the six-month run period of \emph{g14}. Various events such as transfer of polarization from H to D, reversal of H polarization, rotation of the main field direction (causing both H and D to reverse) and an accidental magnet quench are shown \cite{Wei13}.}
    \label{fig:polar}
\end{figure}

The principal function of the IBC is to hold a polarized HD target for experiments in the Jefferson Laboratory Hall B spectrometer, CLAS.  Figure~\ref{fig:polar} depicts the target polarizations of H and D during a portion of the E06-101 (\emph{g14}) experiment and exhibits that the IBC's combination of magnetic field and temperature produces lifetimes for the polarizations which is much longer than the experimental run.  In fact, the cooling performance during this 6 months run was excellant, maintaining a temperature of 50 to 70~mK except for target manipulations, magnetic field manipulations including two ``quench'' incidents, and the electron beam tests.  The two magnetic failures were due to a malfunctioning magnetic power supply and a computer network failure resulting in a drop in liquid helium delivery that uncovered the magnet.  Both were caused by events external to the cryostat.

 Figure~\ref{fig:polar} also illustrates several polarization manipulations for target 22b.  At day 21, following target implantation, the polarizations of H and D are shared so that H goes down from 50\% to 26\% while D rises from 16\% to 26\%. On day 35, the H polarization is flipped with a 90\% efficient RF adiabatic fast passage.  Some loss of D occurs at the same time.  Shortly thereafter, the holding field direction is reversed, reversing both polarization directions, with minimal impact on H polarization magnitude but with some loss of D.  A similar loss occurs again at day 45 when the field is rotated back.  A likely explanation is that, the D polarization lifetime is too short at the low fields used for the H RF flip and available from the transverse saddle coil for the field rotation.

\begin{figure}
    \centering
    \includegraphics[width=\linewidth]{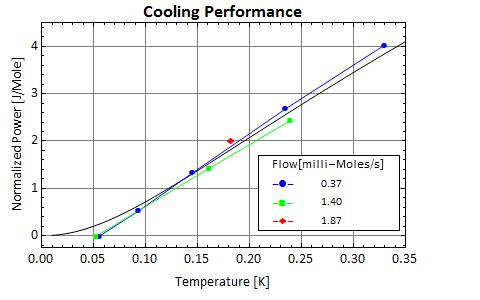}
    \caption{Normalized Cooling Power (heater power $\div$ helium-3 flow rate) versus temperature.   Blue dots are measured with turbo pumping while green squares and red diamonds are measured with Roots pumping.  The colored lines connecting the points are to guide the eye.  The black curve is the theoretical performance for an ideal refrigerator \cite{Rade67}.}
    \label{fig:coolpwr}
\end{figure}

Another measure of the in-beam cryostat's performance is the cooling power, the external heat applied to the mixing chamber to warm to various temperatures.  Measurements of this are plotted in Figure \ref{fig:coolpwr} where the cooling powers have been normalized to the circulation rate. The 1.4 milli-moles/s in Figure \ref{fig:coolpwr} was the maximum flow available during the \emph{g14} run because of a partial block in the return helium-3 line resulting from an air leak during cooldown.  Subsequent measurements determined that flow rates up to 1.9 milli-moles/s could be used although the helium-3 content of the return gas drops to 89\% and the still temperature warms to 860~mK. This indicates the still limits operation at  even higher flow rates.  Nevertheless, the temperatures in Figure \ref{fig:coolpwr} are all consistent with theory \cite{Rade67} and, at the 1.9 milli-moles/s flow rate, a temperature of 250~mK is implied for 5 mW heating, more than satisfying the design goal of 300~mK for electron running.

In summary, the device described here satisfies all the design goals and has proven to be robust and reliable for multiple target manipulations and the long term operation with beam on solid polarized hydrogen-deuteride targets.

\section{Acknowledgements}

The authors wish to thank J. Alston, D. Anderson, C. Apeldoorn, W. Clemens, J. Dail, J. Dickinson, P. Hemler, D. McCay, R. W. Teachey, S. Williams, D. Tilles and the Jefferson Lab Hall-B technical crew for their dedicated assistance during the design, construction and commissioning of the in-beam cryostat described herein.  A. Comer and the Jefferson Lab Target Group provided invaluable assistance in the design and in fabrication of the copper sinters and tubing spirals.  This material is based upon work supported by the U.S. Department of Energy, Office of Science, Office of Nuclear Physics under contract DE-AC05-06OR23177.

\end{document}